\documentclass[preprint,12pt]{elsarticle}

\usepackage{color}

\newcommand{\hefour}{\hspace{1pt}\textsuperscript{4}He}

\newcommand{\qop}[1]{\hat{#1}}
\newcommand{\diff}{{\operatorname{d}}}
\newcommand{\confX}{{\boldsymbol{X}}}
\newcommand{\cs}[1]{\texttt{#1}}

\usepackage{varwidth}
\usepackage{setspace}

\usepackage{calc}

\usepackage{graphicx}

\usepackage{amssymb}
\usepackage{amsmath}

\biboptions{comma,square,sort&compress}

\newcounter{bla}

\journal{Computer Physics Communications}

\usepackage{mathtools}

\usepackage{dcolumn}

\usepackage{booktabs}

\usepackage{colortbl}

\usepackage{rotating}
\usepackage{sidecap}
\usepackage{subfig}

\begin{document}

\begin{frontmatter}



\title{Fast quantum Monte Carlo on a GPU}


\author{Y.~Lutsyshyn\corref{cor1}}
 \ead{email address}
\ead[url]{www.physik.uni-rostock.de/qtmps}
\cortext[cor1]{Corresponding author. Tel.: +49 3814986953 }
\address{Institut f\"ur Physik, Universit\"at Rostock, 18051 Rostock, Germany}

\begin{abstract}

We present a scheme
for the parallelization of quantum Monte Carlo method on graphical processing units,
focusing on variational Monte Carlo simulation of bosonic systems.
We use asynchronous execution schemes with shared memory
persistence, and obtain an excellent
utilization of the accelerator.
The CUDA code is provided along with a package that 
simulates liquid helium-4.
The program was benchmarked on several models of Nvidia GPU, including 
Fermi GTX560 and M2090, and 
the Kepler architecture K20 GPU.
Special optimization was developed for the Kepler
cards, including placement of data structures in the 
register space of the Kepler GPUs.
Kepler-specific optimization is discussed.

\end{abstract}


\end{frontmatter}

\noindent
{\bf Program Summary} \\
\begin{small}
{\em Manuscript Title:} Fast quantum Monte Carlo on a GPU     \\
{\em Authors:} Yaroslav Lutsyshyn                             \\
{\em Program Title:} QL                                       \\
{\em Journal Reference:}                                      \\
{\em Catalogue identifier:}                                   \\
{\em Licensing provisions:} none                              \\
{\em Programming language:} CUDA-C, C, Fortran                \\
{\em Computer:} PC Intel i5, Xeon cluster, GPUs GTX 560Ti, Fermi M2090, Tesla K20  \\
{\em Operating system:} Linux                                 \\
{\em RAM:} Typical execution uses as much RAM as is available on the GPU; usually between 1~GB and 12~GB.
Minimal requirement is 1~MB.                                  \\
{\em Number of processors used:}                              \\
{\em Supplementary material:}                                 
{\em Classification:}  4 Quantum Monte Carlo, 6 CUDA          \\
{\em External routines/libraries:}                            \\
{\em Subprograms used:}                                       \\
{\em Nature of problem:}\\
QL package executes variational Monte Carlo
for liquid helium-4 with Aziz II interaction potential 
and a Jastrow pair product trial wavefunction.
Sampling is performed with a Metropolis
scheme applied to single-particle updates. 
With minimal changes, the package
can be applied to other bosonic fluids,
given a pairwise interaction potential and
a wavefunction in the form of a product of one- and two-body correlation factors. 
   \\
{\em Solution method:}\\
The program is parallelized for execution with Nvidia GPU. 
By design, the
generation of new configurations is performed with shared memory persistence and 
the asynchronous execution allows for the CPU load masking.
   \\
{\em Restrictions:}\\
  Code is limited to variational Monte Carlo.
  Due to the limitation of the shared memory of GPU, only systems under
  2$\,$000 particles can be treated on the Fermi generation cards, and
  up to 10$\,$000 on Kepler cards. 
   \\
{\em Unusual features:}\\
{\em Additional comments:}\\
{\em Running time:}\\
  Because of the statistical nature of Monte Carlo calculations, computations may be chained
  indefinitely to improve statistical accuracy.
  As an example, using the QL package, the energy of a liquid helium system with 1952 atoms can   
  can  be computed to within 1mK per atom in less than 20 minutes.
  This corresponds the relative error of $10^{-4}$. 
  It is unlikely that a higher accuracy may be needed.
   \\
\end{small}



\newpage

\section{Introduction}

Quantum Monte Carlo (QMC) is
an umbrella term for a family of first-principles 
methods for solving quantum many-body problems.
These are the methods of choice 
for describing correlated quantum liquids and solids,
with a wide range of applications \cite{CeperleyQMCReview2010,Ceperley2014-FirstPrinciplesMethodsAPerspectiveFromQuantumMonteCarlo,NeedsQMC2010}.
The methods scale well and 
can be used on systems with hundreds and
even thousands of particles.
QMC is nonetheless computationally expensive,
creating a demand for efficient parallelization
schemes.
In this work we present a parallelization 
of QMC for graphical processing units (GPU).

A GPU processor is built with a number of transistors
which is comparable to that of a modern CPU, but 
allocates more transistors to the arithmetic logical units,
commonly called single processor cores. 
A GPU from the Nvidia Kepler family, for example, hosts over two thousand 
such ``cores''. When properly programmed, GPUs allow 
to execute numerical algorithms with significant acceleration.
However, the large number of GPU cores are created in  expense of
the cache memory and control units. 
The challenge is to supply such a large number
of cores with enough workload and to avoid serialization.

Several groups reported successful attempts of GPU parallelization for quantum Monte Carlo %
\cite{  %
Yunoki2013-AStudyOfParallelizingONGreenFunctionBasedMonteCarloMethodForManyFermionsCoupledWithClassicalDegreesOfFreedom,%
Uejima2011-AccelerationOfAQMMMQMCSimulationUsingGPU,%
Uejima2013-GPGPUforOrbitalFunctionEvaluationWithANewUpdatingScheme,%
Ragavan2013-GPUAccelerationOfTheVariationalMonteCarloMethodForManyBodyPhysics,%
Anderson2007-QuantumMonteCarloOnGraphicalProcessingUnits,%
Ceperley2012-AcceleratingQuantumMonteCarloSimulationsOfRealMaterialsOnGPUClusters%
}%
.
Most of these works relate to QMC methods that 
either operate in the second quantization formalism or use an expansion in finite basis sets
and thus find their bottlenecks in
operations on very large matrices, or even in the generation of pseudorandom number sequences
\cite{Yunoki2013-AStudyOfParallelizingONGreenFunctionBasedMonteCarloMethodForManyFermionsCoupledWithClassicalDegreesOfFreedom,%
Uejima2011-AccelerationOfAQMMMQMCSimulationUsingGPU,%
Uejima2013-GPGPUforOrbitalFunctionEvaluationWithANewUpdatingScheme,%
Ragavan2013-GPUAccelerationOfTheVariationalMonteCarloMethodForManyBodyPhysics}. 
The method that is considered here uses the real-space representation,
and the numerical challenges lie in the calculation of
pair distances, estimation of wavefunctions and expectation values. 
Numerically, QMC in space representation 
is rather different from the above methods.

While the real-space representation for QMC is a wide-spread approach,
GPU parallelization of such methods has only been reported in 
Refs.~%
\cite{Anderson2007-QuantumMonteCarloOnGraphicalProcessingUnits,%
Ceperley2012-AcceleratingQuantumMonteCarloSimulationsOfRealMaterialsOnGPUClusters}.
Anderson, Goddard, and Schr\"oder described their GPU efforts in 2007 \cite{Anderson2007-QuantumMonteCarloOnGraphicalProcessingUnits}.
This early implementation masked the data structures 
as graphical objects and used a GPU to process them. 
Such an approach necessarily had significant limitations.
Nevertheless, Anderson et al.\
achieved acceleration, comparing to a single-core execution, of up to  
a factor of $\times7$.
It was concluded in \cite{Anderson2007-QuantumMonteCarloOnGraphicalProcessingUnits} that 
the Amdahl law \cite{Amdahl-1967} imposed limitations to the achieved acceleration.
That is, the serial load represented a significant bottleneck.
%
Presently, GPU may be programmed with specially designed languages,
most notably with CUDA for Nvidia GPUs. Recently, Elser, Kim, Ceperley, and Shulenburger
reported a CUDA implementation for the QMCPack package \cite{Ceperley2012-AcceleratingQuantumMonteCarloSimulationsOfRealMaterialsOnGPUClusters}. 
Compared to the execution on four cores of a Xeon processor, the GPU-parallelized
code ran up to $\times16$ times faster. 
The works
\cite{Anderson2007-QuantumMonteCarloOnGraphicalProcessingUnits,%
Ceperley2012-AcceleratingQuantumMonteCarloSimulationsOfRealMaterialsOnGPUClusters}
established
certain approaches, in particular, a successful use of the mixed
precision for representing floating point numbers.
It should be noted that while both of Refs.~\cite{Anderson2007-QuantumMonteCarloOnGraphicalProcessingUnits,Ceperley2012-AcceleratingQuantumMonteCarloSimulationsOfRealMaterialsOnGPUClusters} 
considered both variational and diffusion Monte Carlo,
this work is limited to variational Monte Carlo (VMC). 
Also, both of Refs.~\cite{Anderson2007-QuantumMonteCarloOnGraphicalProcessingUnits,Ceperley2012-AcceleratingQuantumMonteCarloSimulationsOfRealMaterialsOnGPUClusters}
treated fermionic wavefunctions for electrons. 
The present work deals with a bosonic quantum liquid, and uses liquid \hefour\ as an example.
It should be noted that the nature of numerical challenges differs significantly 
for the fermionic problems, which require a determinant wavefunction.
Our bosonic wavefunction is a Jastrow product of pair correlation factors.

A good numerical throughput on the GPU became possible only after several enhancements. 
In particular, asynchronous walker writeouts from GPU 
to host memory allowed to limit
reloading of walkers into the shared memory of the GPU. 
As the essential data remains 
loaded into high-level memory of the GPU,
the GPU proceeds generating new configurations without delay.  
For the new Kepler family of Nvidia GPUs, we 
made use of its very large register space, allowing
for better occupancy of the GPU and for the simulation of significantly larger systems.
Asynchronous kernel execution allowed masking of the CPU load 
by the GPU run times, thus partially circumventing the Amdahl limit.
These and several other features are described in detail below, 
and a working minimal code is provided.

\section{Variational Monte Carlo}
For a review of quantum Monte Carlo, readers may refer to 
Refs.~\cite{
LesterBook1994,
CeperleyQMCReview2010,
NeedsQMC2010,
Ceperley2014-FirstPrinciplesMethodsAPerspectiveFromQuantumMonteCarlo,
QMBT2002}.
The variational Monte Carlo relies on the variational principle
which states that for any approximate wavefunction $\psi$ of a system
with Hamiltonian $\qop{H}$,
its ground-state energy $E_0$ is bounded from above by 
\begin{equation}
E_0 \le \frac{\langle \psi | \qop{H} | \psi \rangle}{\langle \psi | \psi \rangle}
.
\label{eq:Ritz}
\end{equation}
Minimization of the r.h.s.\ of Eq.~(\ref{eq:Ritz}) with respect to $\psi$
yields an upper bound for the ground state energy $E_0$.
Optimized trial wavefunction $\psi$ may be taken as the best guess for the true ground state of the system.
Physical properties are extracted by evaluating the
expectation values of their corresponding quantum operators,
\begin{equation}
\langle \qop{A} \rangle = \frac{\langle \psi | \qop{A} | \psi \rangle}{\langle \psi | \psi \rangle}
.
\label{eq:OpExpectation}
\end{equation}
In particular, optimization of the trial wavefunction is the  minimization of $\langle \qop{H} \rangle$.
The optimized variational wavefunction
is also used for importance sampling in 
first-principles many-body calculations, in particular
the diffusion Monte Carlo \cite{Kalos1974-HeliumAtZeroTemperatureWithHardSphereAndOtherForces} and sometimes for the path integral ground state Monte Carlo \cite{Schmidt2000-APathIntegralGroundStateMethod}. 

The trial wavefunction $\psi$ is defined through a set of 
\emph{variational parameters} $\{\alpha_i\}$ 
and the minimization implied by Eq.~(\ref{eq:Ritz}) is 
performed with respect to these parameters.
Using a real space representation of the many-body wavefunction $\psi$,
\begin{equation}
\langle \psi_{ \{\alpha\} } | \qop{A} | \psi_{ \{\alpha\} } \rangle  
=
\int \diff \boldsymbol{r}_1 \cdots \diff \boldsymbol{r}_N 
\,
\psi^\ast_{ \{\alpha\} }(\boldsymbol{r}_1,\dots,\boldsymbol{r}_N)
\,
\qop{A}
\,
\psi_{ \{\alpha\} }(\boldsymbol{r}_1,\dots,\boldsymbol{r}_N),
\label{eq:MainIntegral}
\end{equation}
where $\boldsymbol{r}_1$ through $\boldsymbol{r}_N$ denote coordinates 
of all particles in the system.
In variational Monte Carlo, the multidimensional integral (\ref{eq:MainIntegral}) 
is evaluated via a Monte Carlo integration scheme,
and this constitutes the numerical challenge of the method.

The Monte Carlo integration of Eq.~(\ref{eq:MainIntegral}) is carried out  by 
sampling of the configuration space
\footnote{Additional degrees of freedom are readily accommodated.}%
\begin{equation*}
\boldsymbol{X}=\{\boldsymbol{r}_1,\dots,\boldsymbol{r}_N\}
\end{equation*}
with probability proportional to the 
non-negative density 
$\psi^\ast(\boldsymbol{X})\psi(\boldsymbol{X})$ \cite{%
McMillan,SchiffVerlet-1967, %
Ceperley1978-MonteCarloStudyOfTheGroundStateOfBosonsInteractingWithYukawaPotentials,%
Ceperley1977-MonteCarloSimulationOfAManyFermionStudy%
} %
.
Normalization implies that the
distribution of samples in the configuration space follows
the probability density given by
\begin{equation}
P(\boldsymbol{X})\,\diff \confX 
= \frac{\psi^\ast(\boldsymbol{X})\psi(\boldsymbol{X})}{\langle \psi  |  \psi \rangle}\,\diff \confX.
\label{eq:Probability}
\end{equation}
Averaging over this distribution the \emph{local value} of an operator, defined as
\begin{equation*}
A_L(\confX)=\frac{\qop A \psi(\confX)}{\psi(\confX)}
,
\end{equation*}
yields the integral of Eq.~(\ref{eq:MainIntegral}) and thus the operator's expectation value (\ref{eq:OpExpectation}).

In this work, the distribution of Eq.~(\ref{eq:Probability}) is sampled with the Metropolis scheme \cite{Metropolis-1949}.
The position in the configuration space $\confX$ is transformed to a position $\confX^\prime=D(\confX)$
and the new position is accepted 
with probability equal to the ratio of the probability densities in the transformed and original configurations.
Thus the updated configuration is taken as
\begin{equation}
\confX^\text{updated} \leftarrow
\begin{cases}
\confX^\prime,&\text{if}\,\,\,\,\,\displaystyle\frac{\left|\psi(\confX^\prime)\right|^2}{\left|\psi(\confX)\right|^2} > \xi \\
\confX,&\text{otherwise}\phantom{\displaystyle\frac{\left|\psi(\confX^\prime)\right|^2}{\left|\psi(\confX)\right|^2}}
,
\end{cases}
\label{eq:Decision}
\end{equation}
where $\xi$ is a pseudorandom number uniformly distributed on $[0,1]$.
The Markov chain produced with the acceptance criterion
(\ref{eq:Decision})
provides configurations distributed according to the desired probability density (\ref{eq:Probability}).

We aim to describe the bulk properties of liquid helium-4. 
With a very good accuracy, helium atoms 
may be considered particle-like with a known pairwise interaction.
The Hamiltonian is given by
\begin{equation*}
\qop{H}=-\frac{\hbar^2}{2m_{He}}\sum_i^N \nabla_i^2 + \sum_{i<j}^N V(r_{ij})
,
\end{equation*}
where $m_{He}$ is the mass of the \hefour\ atom, $N$ is the number of atoms in the
system, and $r_{ij}$ is the distance between particles $i$ and $j$. 
For the pairwise interaction $V(r)$, we will use 
HFD-B(HE) (known as Aziz-II)
interatomic potential  \cite{AzizII}.

The form of the wavefunction which we implemented
is a widely used Jastrow product of pair correlation
factors \cite{Jastrow},
\begin{equation}
\psi(\confX)=
\prod_{i<j}^{N}\operatorname{e}^{u(r_{ij})} 
=\operatorname{e}^{\,\sum_{i<j}u(r_{ij})}.
\label{eq:Jastrow}
\end{equation}
The choice of wavefunction fully defines the probability 
density for the Metropolis sampling in variational Monte Carlo, according to Eq.~(\ref{eq:Decision}).
We have implemented the parallelization for single-particle updates, where only 
one atom is displaced at a time to form the new configuration $\confX^\prime$.
The transformed (trial) configuration $\confX^\prime$ is given by
\begin{equation*}
\confX^\prime = \{\boldsymbol{r}_1,\dots,\boldsymbol{r}_{s-1},\boldsymbol{r}_s+\boldsymbol{\delta},\boldsymbol{r}_{s+1},\dots,\boldsymbol{r}_N\}
,
\end{equation*}
where $s$ is a randomly selected particle index and 
$\boldsymbol{\delta}$ is a randomly selected displacement.
We use Gaussian-distributed displacements.
Single-particle updates result in 
autocorrelation 
times in the generated Markov chain that only scale as the first order in the number of particles.
Such updates are suitable for VMC sampling of quantum liquids with large number of atoms.
For single-particle updates of a system given by~(\ref{eq:Jastrow}), the acceptance probability
in a  Metropolis step~(\ref{eq:Decision}) becomes
\begin{equation}
\frac{\left|\psi(\confX^\prime)\right|^2}{\left|\psi(\confX)\right|^2}
=
\exp\left\{2 \sum_{i\ne s}\left[   u(|\boldsymbol{r}_s+\boldsymbol{\delta}-\boldsymbol{r}_i|)-u(|\boldsymbol{r}_s-\boldsymbol{r}_i|)  \right]  \right\}
.
\label{eq:JastrowUpdate}
\end{equation}
We use a symmetrized form for $u(r)$,
\begin{equation}
u(r)= \begin{cases}  \displaystyle
\frac{f(r)+f(L-r)}{2f(L/2)} &,\,\, r\le L/2\\
1 &,\,\, r>L/2
.
\end{cases}
\label{eq:uFunction}
\end{equation}
The symmetrization length $L$ is usually chosen as the smallest box dimension.
Symmetrization results in a continuous first derivative of 
$\psi(\confX)$, removes tail corrections to the kinetic energy and
allows to use alternative estimators for the this energy.
The pair-correlation factor $f(r)$ is chosen in the Schiff-Verlet form \cite{SchiffVerlet-1967}
\begin{equation}
f(r)=-\frac{1}{2}\left(\frac{b}{r}\right)^5.\
\label{eq:fifthPower}
\end{equation}
This function prevents the overlap of atomic cores.
The power in $r^{-5}$ is
dictated by the behavior of helium-helium interaction potential 
at short distances and is close to optimal for the Aziz-II potential. 
The parameter $b$ is the only variational parameter in this wavefunction.

Much more complicated forms of the trial wavefunctions exist
for strongly correlated quantum liquids and solids
\cite{%
Reatto1979-SpatialCorrelationsAndElementaryExcitationsInManyBodySystems,%
Loubeyre1987-ThreeBodyExchangeInteractionInDenseHelium,%
Lutsyshyn2010-PropertiesOfVacancyFormationInHcp4HeCrystalsAtZeroTemperatureAndFixedPressure,%
Schmidt1980-VariationalMonteCarloCalculationsOfLiquid4HeWithThreeBodyCorrelations%
}.
Even for the Jastrow form of Eq.~(\ref{eq:Jastrow}), 
the pair factor $u(r)$ can be considerably more complex 
and computationally expensive. Triple-correlation
forms are also used \cite{Schmidt1980-VariationalMonteCarloCalculationsOfLiquid4HeWithThreeBodyCorrelations}. 
However, additional arithmetics increase the parallel fraction
of the program and only improves the parallelization outcome.
The simple choice of the wavefunction is 
made with the purpose to demonstrate the strength
of the parallelization scheme.
  
\section{GPU features and architecture\label{sec:CUDA}}
In this section, we introduce the relevant terminology, and review a number of GPU features that were taken advantage of in our program. 


Single computing cores of the GPU
are grouped into units called streaming multiprocessors (SM or SMX for Kepler architecture).
Each SM contains additionally several special function units,
a large amount of register space and a high-level user-managed cache called \emph{shared memory}. 
A GPU unit typically includes between one and twelve GB 
of an on-board high-speed operating memory accessible to all SMs. This memory
is called \emph{global}.

A routine commanded by a CPU to be executed on a GPU is called a \emph{kernel}. 
A kernel is executed by a  
certain number of \emph{thread blocks}. 
Each thread block is resident on a
multiprocessor and the threads within a block are easily synchronized. 
Communication and synchronization between different thread blocks is, however, 
slow and in principle not guaranteed by the language standard.
To compute quantities to desired precision, it is productive to generate several independent 
Markov chains, each thread block working on its own   walker.
Only for very large systems,
it becomes advantageous to use all multiprocessors simultaneously to work on a single Markov
chain. 
Advancing a single walker is also preferable for systems with very long equilibration times.
Such a cooperating approach, with synchronization written along the lines of Ref.~\cite{Feng2010-ToGPUSynchronizeOrNotGPUSynchronize}, 
has been successfully tried out 
but is not included here.
Using multiple walkers is also more compatible with other 
QMC methods.

By design, each multiprocessor is capable of maintaining a very large 
number of threads with an intact register space for each thread, 
and can rapidly switch between the threads. 
Quite unlike for CPU, the GPU masks memory latency and 
the latency of mathematical units by switching between threads in a thread block and 
executing the instructions for the threads that are ready to proceed.
It is thus favorable to run a very large number of threads on each SM,
much larger than the actual number of single cores. 
We take advantage of this by partitioning
the calculation of interparticle distances to different threads,
each thread only computing several distances.
The number of threads in a threadblock does not have to equal the number of particles.

Successive chunks of 32 threads, called \emph{warps}, simultaneously execute the same 
instruction sequence but with their own data. Conditional branching within warps is possible 
but results in serialization. Thus all threads in a given
warp are automatically synchronized. 
The \emph{warp synchronicity} 
is heavily exploited in the QL package, especially when performing
parallel reduction such as for the sum of Eq.~(\ref{eq:JastrowUpdate}).
This leads to a noticeable improvement in the execution times.
The functionality can be switched by a preprocessor macro, as warp synchronicity is not 
strictly a part of the CUDA standard.
Warp synchronicity is exercised on the thread level 
and it can be used even when the number of particles is not a multiple of  
the warp size.

GPU kernels may write directly to the host memory via the front-side bus.  
This process does not require terminating the kernel, allowing for shared memory persistence.
All the caches and SM shared memory content and even registers may be kept intact. 
The data transfer to the host memory
is also asynchronous with respect to the kernel's execution. 
The kernel proceeds to the
next operation after posting the memory command, even though the memory transaction completes
thousands of cycles later. 
We  take advantage of this feature by 
keeping the walkers in the shared memory, and regularly ``writing out'' a 
copy to the central RAM. The SM proceeds to the next 
calculation immediately, and does not need to reload the coordinates.
This feature was essential for achieving the reported performance.

Once a kernel is launched, control returns to the CPU, without waiting for the GPU task to finish.
This capability is exploited to run the GPU and CPU parts of the code simultaneously.
These are referred here as the GPU-side and CPU-side. 
The GPU performs the heavy number crunching; the code on CPU at the same time processes generated configurations
and prepares raw random numbers for the next chunk of calculations. So long as the execution times of the CPU-side are 
shorter than those of the GPU, the load of CPU computations is almost completely masked. 
This is further discussed in Section~\ref{sec:LoadMasking}.

\section{Load masking, optimal execution and the modified Amdahl law \label{sec:LoadMasking}}


\subsection{Optimal execution}

Variational Monte Carlo calculations can be seen as consisting
of two interleaved tasks: generating the Markov chains to produce new configurations (according to the goal distribution)
and then computing various observables on these configurations.
We refer to the former as the generation steps, and to the latter as the analysis steps.
A number $N$ of single-particle updates (sometimes called \emph{microupdates}), where $N$ is the number of particles, constitute 
a \emph{macroupdate} and separate configurations
$\boldsymbol{X}_j$ and $\boldsymbol{X}_{j+1}$ in our notation.
The cost of one microupdate includes
computing $N$ interparticle distances for the sum of Eq.~(\ref{eq:JastrowUpdate}).
Thus the cost of performing one macroupdate is $\mathcal{O}(N^2)$, 
the same as that of a typical analysis step, but possibly with quite a different coefficient.
The consecutive configurations are correlated,
and one needs to decide the optimal number of macroupdates
to be performed between the analysis steps.
Optimal conditions result in the smallest statistical errors 
given the available computer time.

For a stationary  walk, 
the autocorrelation function $R(k)$ of the sequence can be 
defined as 
\[
R(k)=\frac
{
\sum_i \left( A_i - \langle A \rangle \right)\left( A_{i+k} - \langle A \rangle \right) 
}
{
\sum_i \left( A_i - \langle A \rangle \right)  ^2 
}
,
\]
where $A_i$ is the sequence of calculated observable values,
and $\langle A \rangle$ is its mean.
We further define the autocorrelation time $m_c$ of a sequence  
as the sum of the autocorrelation function \cite{LandauBinderBook},
\[
m_c = \sum_{k=1}^\infty R(k).
\]
For an uncorrelated sequence, $R(k)=\delta_{k,0}$,
and the autocorrelation time $m_c$ defined in this way is equal to zero%
\footnote{The use of word ``time'' here should be understood to represent
the position along the Markov chain, and not actual physical 
dynamics, which is absent from Monte Carlo simulations.
}%
.
While the above definitions can be used to perform the autocorrelation 
analysis directly \cite{Geyer1992-PracticalMarkovChainMonteCarlo,ThompspnThesis2011},
the autocorrelation time can also be extracted by reblocking techniques 
\cite{%
Flyvbjerg1989-ErrorEstimatesOnAveragesOfCorrelatedData,%
NeedsQMC2010%
}.

Given $m_c$ is the autocorrelation time between successive configurations,
and that configurations are only analyzed every $m$ macroupdates (that is, one 
only performs analysis on configurations $\boldsymbol{X}_j$, $\boldsymbol{X}_{j+m}$, $\boldsymbol{X}_{j+2m}$, \dots), 
the true uncertainty in the mean of a computed quantity 
is\footnotemark
\begin{equation}
\sigma=\frac{\sigma_0}{\sqrt{n_\text{samples}}}\sqrt{1+\frac{2 m_c}{m}}
,
\label{eq:CorrelatedError}
\end{equation}
where $\sigma_0^2$ is the variance of the 
distribution of this quantity and $n_\text{samples}$ 
is the number of analysis instances 
\cite{LandauBinderBook}.
This expression is not modified in the important case when several
independent Markov chains are used. Given that $n_\text{samples}$ 
is the total number of the samples collected in all chains and the number of the
chains is $W$, the number of samples in each chain is $n_\text{samples}/W$ and 
the resulting uncertainty is given by
\begin{equation*}
\sigma=\frac{1}{\sqrt{W}}\frac{\sigma_0}{\sqrt{n_\text{samples}/W}}\sqrt{1+\frac{2 m_c}{m}}
=
\frac{\sigma_0}{\sqrt{n_\text{samples}}}\sqrt{1+\frac{2 m_c}{m}}
.
\end{equation*}
This reasoning applies so long as the number of samples
in each chain is large, $n_\text{samples}/W\gg 1$.
Thus, regardless of the value of $m_c$, progressing several 
independent chains is as good as progressing a single but correspondingly longer chain.
This equally applies to the case of very strong autocorrelation.
A significant caveat, however, is that the chains have to be independent. 
This may become a  difficulty for systems with very large initial 
equilibration times. Such a situation is discussed further below.

\footnotetext{
For the following analysis, we assume that the 
central limit theorem (CLT) applies strictly to the computed quantities. 
This is the case for our system. However, readers should
be aware that weaker forms of CLT may arise, especially 
for nodal wavefunctions \cite{Trail2008-HeavyTailedRandomErrorInQuantumMonteCarlo,Trail2010-OptimumAndEfficientSamplingForVariationalQuantumMonteCarlo}. 
}

For fixed computer time, $n_\text{samples}$ is inversely proportional
to the time between analyses $t_\text{step}$. 
Let $t_\text{A}$ be the time necessary for analysis,
 and $t_\text{G}$ the generation time necessary to produce one 
macroupdate.
If both the generation and analysis are performed on the same unit,
\begin{equation}
t_\text{step}=t_\text{A}+m t_\text{G},
\label{eq:serialtime}
\end{equation}
and the optimization of Eq.~(\ref{eq:CorrelatedError})
gives 
\begin{equation}
m=\sqrt{2m_c\frac{t_\text{A}}{t_\text{G}}}=\sqrt{2 m_c \alpha}.
\label{eq:optimalCPU}
\end{equation}
Above we introduced $\alpha$,
the ratio of time needed for analysis and macroupdate generation, 
\begin{equation*}
\alpha=t_\text{A}/t_\text{G}.
\end{equation*}
Consider now the asynchronous generation, in 
which the GPU generates new macroupdates, and
the CPU performs analysis on previously prepared configurations 
at the same time. 
The asynchronous time per analysis instance is 
\begin{equation}
t_\text{step}=\max(m t_\text{G}, t_\text{A}).
\label{eq:maskedtime}
\end{equation}
The optimization of Eq.~(\ref{eq:CorrelatedError})
yields, unsurprisingly, 
\begin{equation}
m=\frac{t_\text{A}}{t_\text{G}}=\alpha
,
\label{eq:optimalGPU}
\end{equation}
that is, both CPU and GPU are used at all times. The time that 
GPU takes to generate $m$ macroupdates is $m t_\text{G}$. Meanwhile, 
the CPU side has just enough time to perform the analysis, since Eq.~(\ref{eq:optimalGPU}) requires $t_\text{A} = m t_\text{G}$.

\subsection{Selecting the optimal analysis regime}

We can now estimate the penalty that one pays by using 
a non-optimal number of the analysis instances.
The computer time necessary to reach a given error level $\sigma^\ast$ is given 
by Eq.~(\ref{eq:CorrelatedError}) as
\[
t=n_\text{samples} t_\text{step} =
\left(\frac{\sigma_0}{\sigma^\ast}\right)^2  \left(1+\frac{2m_c}{m}\right) t_\text{step}.
\]
It is a common practice to perform analysis after every macroupdate, corresponding to $m=1$.
We begin with the case when both the generation and analysis are
performed sequentially on the same unit%
\footnote{This applies both to a fully CPU-bound calculation
and to the simulations with the observables being computed
on the GPU along with the Markov chain generation.}.
Using Eqs.~\mbox{(\ref{eq:CorrelatedError}--\ref{eq:optimalCPU})},
one can find the ratio $P_\text{seq}$ of computer times $t_\text{seq}$ necessary 
to reach a target uncertainty level when either $m=1$ or $m$ is
selected optimally as given in Eq.~(\ref{eq:optimalCPU}),
\begin{equation}
P_\text{seq}(m=1)=
\frac{t_\text{seq}(m=1)}{t_\text{seq}(m=\sqrt{2m_c\alpha})}
=
\frac
{
(\alpha+1)(1+2m_c)
}
{
\alpha\left(1+\sqrt{\frac{2m_c}{\alpha}}\right)^2
}
.
\label{eq:penaltyseq}
\end{equation}
The largest penalty occurs for strong autocorrelation, 
when the penalty ratio reaches the value $P_\text{seq}=\alpha+1$.
If the cost of computing the observable is similar to the cost 
of performing the macroupdate, 
then
we can expect that $\alpha$ is of the order of unity 
since the calculations are performed on the same unit.
The penalty of performing analysis on every step is, therefore, not prohibitive.
On the contrary, the value of $\alpha$ for the GPU-accelerated Markov 
walk with observables computed on the CPU is expected to be considerably 
large. It is closely related to the achieved overall acceleration.
In this case, the ratio $P_\text{mask}$ of computer time 
for $m=1$ and optimal $m$ given by Eq.~(\ref{eq:optimalGPU}) is equal to
\[
P_\text{mask}(m=1)=
\frac{t_\text{mask}(m=1)}{t_\text{mask}(m=\alpha)}=
\frac
{
\alpha(1+2m_c)
}
{
\alpha+2m_c
}
.
\]
This ratio cannot exceed $P_\text{mask}=\alpha$, but in this case the penalty 
can be large. It is caused by the unnecessary idling of the accelerator.

Similar considerations apply when one chooses to perform the analysis some every $m$ steps. 
In this case, the penalty ratio for the sequential case is bounded by 
\[1 \le P_\text{seq}(m) < 1+\max\left(\frac{\alpha}{m},\frac{m-1}{\alpha+1}\right).\]
For masked execution with the GPU, the penalty is bounded by 
\[1 \le  P_\text{mask}(m)  < \max\left(\frac{\alpha}{m},\frac{m}{\alpha}\right).\]
Notice that the upper bounds do not contain the autocorrelation time $m_c$.
It is therefore possible to determine a ``good'' value for $m$ without performing the autocorrelation analysis.
Instead, one can determine $\alpha$ from simple benchmarking,
and select $m \approx \alpha$. For the sequential execution, 
the penalty is at most twofold: $1 \le  P_\text{seq}(m=\alpha) < 2$.
The upper limit is only reached for very strong autocorrelation, when $m_c\gg\alpha$.
For the CPU+GPU execution, it is in any case optimal to choose $m=\alpha$, see Eq.~(\ref{eq:optimalGPU}).
In practice, optimal $m$ is often determined empirically through a series of benchmarks.
In this case, using $m\approx\alpha$ is a good starting point.

When the autocorrelation time $m_c$ is large,
the analysis needs to be performed infrequently 
and can be done on the CPU-side.
The cost of analysis will be masked by the time that that the 
GPU needs to spend creating new configurations.
However, if the autocorrelation time $m_c$ is small,
the analysis should be performed frequently, and the 
the CPU-side becomes the bottleneck. 
If one needs to minimize the errorbars for such an observable, 
one should consider either moving the calculation of the 
observable to the GPU, or accelerating their computation on the CPU.

\subsection{Modified Amdahl law \label{sec:Amdahl}}

\begin{figure}
\begin{center}
\includegraphics[width=1\columnwidth]{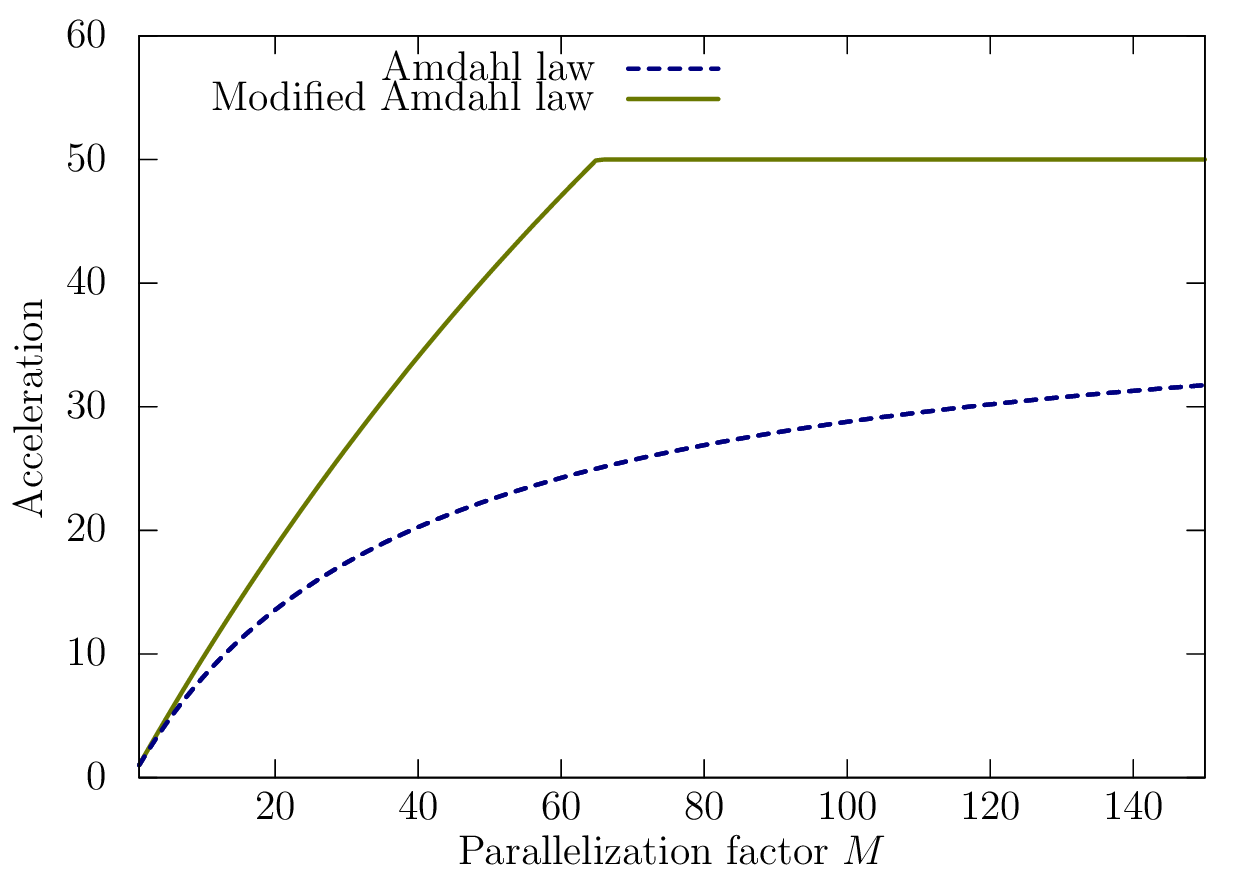} %
\caption{
\label{fig:Amdahl}
(Color online)
Illustration of the modified Amdahl law.
Here, we used $R=2\%$ for the computational fraction remaining on the host, and $S=0.5\%$ for the serial
fraction of the GPU kernel. Horizontal axis shows 
the parallelization factor $M$, as explained in the text. 
The vertical axis shows achievable acceleration ratios. 
The dashed blue line corresponds to the ``traditional'' Amdahl law, that is, the acceleration achievable via synchronous execution as given by Eq.~(\ref{eq:standardAmdahl}). The solid green line 
shows the Amdahl law which results from concurrent (asynchronous) execution and load masking, as shown in Eq.~(\ref{eq:modifiedAmdahlFull}).
While the maximal acceleration values are quite close ($\times40$ and $\times 50$),
the asynchronous execution reaches its limit significantly faster. 
}
\end{center}
\end{figure}

With increased parallelization, even a small serial fraction in a program 
inevitably becomes the main computational bottleneck. This observation is known as the Amdahl law \cite{Amdahl-1967}.
The serial fraction arises not only from parts of the algorithm that were not parallelized, but also
from the communication and synchronization between parallel workers. Given that $S$ is the serial and $P$ the parallel 
fractions of the program ($S+P=1$), and $M$ is the number of parallel threads or processes, the achievable acceleration $A$ is 
given by
\[
A=\frac{S+P}{S+P/M}=\frac{1}{S+(1-S)/M}<\frac{1}{S}.
\]
For a program with the GPU acceleration, all work that remains on the CPU-side
appears as the serial fraction, since it is not parallelized for the accelerator.
The same consideration applies to the communication between the CPU and GPU. 
It is thus often the case that the CPU-side of the program becomes the 
bottleneck of the GPU parallelization, since $M\gg 1$ and the Amdahl limit regime is quickly reached.

The load masking leads to a modified form 
of the Amdahl law. 
Suppose that $R_\text{CPU}$ is the code fraction that remains on the CPU\footnotemark, 
and the serial and parallelizable parts that are
moved to the GPU are $S_\text{GPU}$ and $P_\text{GPU}$.
Thus 
\[
R_\text{CPU}+S_\text{GPU}+P_\text{GPU}=1.
\]
There is an additional overhead associated with
launching the parallel process; it is omitted here for clarity.
\footnotetext{CPU side of the program need not be                                           %
serial. It may, for example, use OpenMP acceleration                                        %
to utilize all available CPU cores.                                                         %
In this case $R_\text{CPU}$ corresponds to the combined                                     %
time of the CPU execution, $R_\text{CPU}=S_\text{CPU}+D^{-1}P_\text{CPU}$, where $D$ is the %
number of CPU threads.}                                                                     %
If the parallel portion of the GPU code may be accelerated by a factor $M$, the resulting speed-up is
\begin{equation}
A_\text{mask}(M)=\frac
{1}
{\max(R_\text{CPU},S_\text{GPU}+M^{-1}P_\text{GPU})}
.
\label{eq:modifiedAmdahlFull}
\end{equation}
For a GPU, the parallelization factor $M$ is related to the number
of available cores, but also strongly depends on a number of other factors.
In the case of a complete CPU-side masking, acceleration reduces to
\begin{equation}
A_\text{mask}(M)=\frac
{1}
{S_\text{GPU}+M^{-1}P_\text{GPU}}
=
\frac
{M}
{M S_\text{GPU}+P_\text{GPU}}
,
\label{eq:modifiedAmdahlMasked}
\end{equation}
with the CPU fraction $R_\text{CPU}$ completely removed from the equation. 
Because the serial component of Markov chain 
generation is much smaller than the serial fraction of the entire program, we have 
\mbox{$S_\text{GPU}\ll R_\text{CPU}\ll1$}.
Having isolated the serial GPU fraction $S_\text{GPU}$, 
we may focus on minimizing it by improving the program.
In the absence of load masking, the acceleration is limited 
by $S_\text{GPU} + R_\text{CPU} \approx R_\text{CPU}$, 
and the CPU-side execution quickly becomes the bottleneck.

The maximum acceleration $A_\text{mask}$ that can be achieved under 
the modified version of the Amdahl law of Eq.~(\ref{eq:modifiedAmdahlFull}) 
is given by $R^{-1}$.
If the execution of the GPU-side was synchronous with the CPU-side, the standard
Amdahl law would apply, with the acceleration given by
\begin{equation}
A_\text{syn}(M)=\frac
{1}
{R_\text{CPU}+S_\text{GPU}+M^{-1}P_\text{GPU}}
.
\label{eq:standardAmdahl}
\end{equation}
The maximum acceleration in this case is given by $(R+S)^{-1}\lesssim R^{-1}$, 
that is, not much different from the limit of $A_\text{mask}$.
Moreover, it holds that $A_\text{mask}/A_\text{syn}\le 2$, by itself not a remarkable improvement.
It simply reflects the fact that one has two devices to use for 
computation: the GPU and the CPU itself.
What is special about the structure of Eq.~(\ref{eq:modifiedAmdahlFull}) 
is the nature in which $A_\text{mask}$ approaches the maximal value. 
The difference is illustrated in Fig.~\ref{fig:Amdahl}.
For the purpose of illustration, we have chosen $R=0.02$ and $S=0.005$.
The masked execution approaches its maximal without the slow tapering characteristic 
of the traditional Amdahl law. 
Thus, achieving the theoretical 
limit is possible for $A_\text{mask}$, and the smaller required parallelization factor $M$
means that the parallelization overhead needs not grow into a bottleneck of its own.

\subsection{Additional considerations}

Even when the GPU performs both 
the generation and analysis, load masking
improves the performance. For example, 
random numbers may be prepared by the CPU 
for the next block while the GPU is busy, as described below.
Memory transactions through the system bus
are also at least partially masked.
An important consequence is that
other, less computationally intensive tasks and analysis may remain on the CPU without any performance penalty.

The set of coordinates and other data representing each Markov chain is commonly
referred to as a ``walker''.
Optimal execution of the GPU code requires 
a careful selection of the number of walkers (and thus thread blocks). 
The number of walkers should be maximal 
such that all thread blocks still remain simultaneously resident on the GPU. 
For smaller systems, number of threads equal to the number of particles is optimal.
However, reducing the number of threads has the benefit
of smaller relative loop overheads (each thread serves several interparticle distances),
reduced required register space and smaller temporary arrays (allowing for a larger number of resident thread blocks),
and a reduced cost of parallel reduction schemes. 
For this reason, large systems are best computed with the number of threads which is several times smaller than the
number of particles.
To make use of warp synchronicity, it is beneficial to use the number 
of threads such that it is a multiple of the warp size, equal to 32. 
Once the number of threads has been decided, the maximum number of 
resident blocks can be determined from the compute capability of the 
device, which determines maximum number of total resident threads on a multiprocessor. 
The optimal number of blocks may be below this value due to memory restrictions.
Finally, to save time on copying the random numbers to the GPU, it is best
to use the largest number of steps which is allowed by the size 
of the GPU global memory.

Special care must be taken when simulating systems with large equilibration times.
This effectively means that the simulation starts with large autocorrelation time (this interval is called ``equilibration''; observables should not be collected during this initial stage).
The equilibration finishes when the autocorrelation time decreases to its ``normal'' value, and afterwards the observables fluctuate about their average values. 
In this case, it is optimal to first progress a smaller number of walkers (or just a single walker) 
to pass the equilibration stage.
The equilibrated walkers are then replicated
and the simulation continues with the number of walkers that is optimal for the hardware.
For the distributed package, however, there is little speed gain
when using fewer walkers than the number of multiprocessors on the GPU. 
Thus for the equilibration stage, it is optimal to use 
the number of  walkers equal to the the number of multiprocessors.

\section{Implementation \label{sec:Implementation}}

\subsection{Computational workflow}

In this section we discuss in greater detail the workflow of the program. 
As described above, the part of the code that executes on the CPU is
referred to as the CPU-side, and the GPU-side refers to execution on the graphical card.
``Global memory'' is the term used for the RAM memory of the GPU, 
and ``host memory'' refers to the main RAM of the computer.

Upon starting of the program, input files provide starting walker coordinates,
box dimension, random sequence seed and execution parameters.
The CPU-side allocates page-locked 
arrays in host memory 
and the necessary arrays in global memory of the GPU.  
The former are used to pass walker coordinates to and
from the GPU, and the latter are used to supply the GPU with 
random numbers.

For statistical analysis of Monte Carlo results, 
the execution is broken into blocks. 
Values averaged during a block are written to a file and values from successive blocks may
be later averaged and analyzed for correlation 
and standard deviation values \cite{Flyvbjerg1989-ErrorEstimatesOnAveragesOfCorrelatedData}.
Walker coordinates and random seeds are also written to files at the end of each block.
Thus every block-end provides a natural restore point.

An instance at which the GPU writes walker coordinates into
page-locked host memory is here called a \emph{writeout}.
Writeout also includes values such as the acceptance ratio,
computed potential and kinetic energies.
The writeouts are performed by the GPU code within the main computational 
kernel. This allows to achieve shared memory persistence.
It is the persistence that, in our case, contributes most significantly to a good GPU utilization. 
Additional benefit of the writeouts is masking of the CPU load 
as described in Section~\ref{sec:LoadMasking}.
Each writeout occurs into a unique separate space in the host memory. 
In the beginning of
a block, the program writes computationally
impossible values into the writeout arrays.
For example, negative values for particle coordinates. 
After starting the GPU kernel, the CPU-side 
loops over the memory space until the area
to which a given generation of walkers is destined has been written to. 
Freshly arrived data is then taken for processing. The behavior is
assisted by forbidding host caching of the writeout arrays through
the \texttt{cudaHostAllocWriteCombined} flag to the \texttt{cudaHostAlloc} command. 
While it is not guaranteed that the writeout 
data will even begin to arrive before the kernel finishes, we
have tested this behavior on a range of systems and 
always observed properly completed, uncorrupted asynchronous writeouts. 
Because each writeout occurs to its reserved part of memory,
\emph{in the worst-case scenario the data will be received after the 
kernel finishes, preserving program correctness}. There is a build-in check in the program which warns
the users if the kernel in fact executes synchronously%
\footnote{Asynchronous execution is automatically disabled, for example, in case of certain profiling flags to the compiler. 
Synchronicity may also be forced with a preprocessor variable.}.

\begin{sidewaysfigure*}
\includegraphics[angle=270,width=\textwidth]{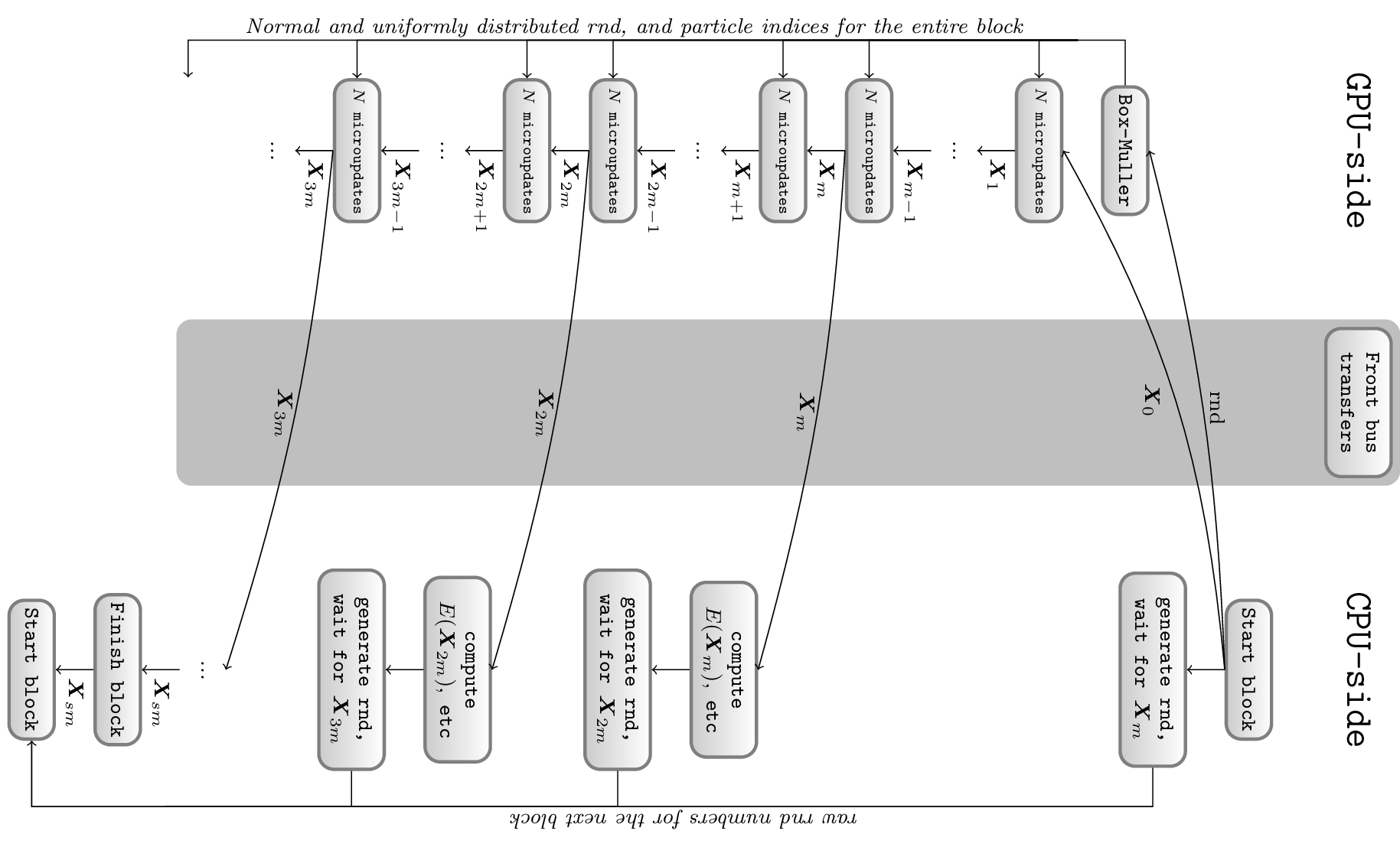} %
\caption{\label{fig:Workflow}
Block-diagram representation of the workflow. 
In this case, the CPU-side generates raw random
numbers and sends them to the GPU-side for Box-Muller processing.
Once the GPU obtains the starting walker configurations, it 
starts the Markov chain. 
Every $m$ macroupdates, the GPU sends the configurations back to the CPU for processing. 
Meanwhile, the GPU continues the generation. 
When the CPU-side is not occupied, it generates
some raw random numbers to be used in the next block. A typical 
calculation consists of a considerable number of such blocks. 
Each block consists of an uninterrupted kernel, with configurations remaining in the shared memory.
The size of a block is limited by the GPU global memory, typically at $2^{28}=268\cdot10^6$ microupdates.}
\end{sidewaysfigure*}

The execution of a block is outlined in Fig.~\ref{fig:Workflow}.
The CPU-side copies random numbers to the GPU,
launches a GPU kernel for the Box-Muller transform, and after that, the main computation kernel.
The Box-Muller transform is used to produce the Gaussian-distributed particle displacements.
Once launched, the kernel loads walker coordinates from page-locked
host memory, and begins to produce the Metropolis updates.
Each thread block serves one walker, that is, one Markov chain of configurations.
For each microupdate, the thread block loads the random particle index $j$, 
the ``dice'' variable $\xi$, and three displacement coordinates $\boldsymbol\delta_j$ from the 
GPU global memory.%
\footnote{Notice that the decision variable is effectively prefetched as it is called at the beginning of 
the microupdate along with several other variables. The value arrives before it is needed at the end of the microupdate.
Such a small difference may account for several percent of performance.}
Each thread computes distances from the updated particle to one or several other particles, 
and computes $u(|\boldsymbol{r}_{j}-\boldsymbol{r}_{i}+\boldsymbol{\delta}_{j}|)-u(|\boldsymbol{r}_{j}-\boldsymbol{r}_{i}|)$ (see Eq.~(\ref{eq:JastrowUpdate})).
Periodic boundary conditions are applied to all interparticle distances.
After this, a parallel reduction scheme is invoked 
and the sum from all threads in a thread block is accumulated
in the zeroth element of the array \texttt{thread\textunderscore sum}.
Zeroth thread makes the update decision according to Eq.~(\ref{eq:Decision}),
\begin{equation*}
d\leftarrow\left\{\exp\left[2\sum_{i\ne j} u(|\boldsymbol{r}_{j}-\boldsymbol{r}_{i}+\boldsymbol{\delta}_{j}|)-u(|\boldsymbol{r}_{j}-\boldsymbol{r}_{i}|) \right]>\xi\right\},
\end{equation*}
where the left-value is an integer variable and 
the logical right-value is evaluated as $1$ when
true and $0$ otherwise.
Update happens non-conditionally  as
\begin{equation*}
\boldsymbol{r}_j\leftarrow\boldsymbol{r}_j+ \boldsymbol{\delta}_j  d.
\end{equation*}
After predetermined number of updates, 
threads perform a writeout of walker coordinates
and other necessary data. All threads participate
in the memory command, with  the data sequence ordered to ensue 
coalesced memory access.

\subsection{Random numbers}

The generation of the (pseudo)random numbers
is separated from the Markov chain generator. 
An entire sequence of ``raw'' pseudorandom numbers (uniformly distributed on the unit interval)
that is necessary for the computational block is generated and placed in the GPU global memory.
The sequence is indicated in Fig.~\ref{fig:Workflow}.
This approach provided  three advantages:
(i)~Pseudorandom (and truly random) number generators may be readily interchanged, making it 
easy to test new generators, especially those that work from the GPU. 
(ii)~Generating a large quantity of random
numbers at once proved significantly faster than producing 
them one-by-one as the need arises.
The gain in speed overcomes the necessary limitation from the bandwidth of RAM
because of the resulting favorable memory access patterns.
(iii)~Numbers may be generated for the next block 
by the CPU while the CPU-side is waiting
on the GPU kernel to finish. This is, in
effect, a masking of CPU load with respect to the pseudorandom number generation, 
as described in Section~\ref{sec:LoadMasking}.

The number of operations
necessary to perform a macroupdate scales
as $\mathcal{O}(N^2)$, while the number
of necessary pseudorandom numbers is $5N$.
Thus producing sufficient amount of random 
numbers is a challenge only for small systems.
In practice, generating raw random numbers
became a bottleneck only for systems with fewer than 256 particles.
The program has a remarkable appetite for the random numbers 
when computing small systems.
For example, in the case of a 64-particle system with 
128 walkers on a Tesla M2090 card, the GPU
consumed over 100 million random numbers per second, rate limited 
by the generator. For such cases, one may opt
for GPU-parallel generators, such as those provided  
by the CURAND library \cite{curandNvidiaURL}.

Pre-generation of the entire set of pseudorandom
numbers makes it convenient to use non-standard sources of uncorrelated sequences.
For instance, we successfully tested our results against the AMU sequence of
true random numbers produced from vacuum fluctuation
measurements \cite{AMUrndGenerator2011-RealTimeDemonstrationOfHighBitrateQuantumRandomNumberGenerationWithCoherentLaserLight}.

\subsection{Register-heavy routines for Kepler architecture}

The Kepler K20 is the latest generation 
GPU from Nvidia. The Kepler chipset sports 
a considerably increased number of computing cores, 
but a reduced number of streaming multiprocessors (called SMX on Kepler).
The amount of shared memory per multiprocessor is the same as in the previous, Fermi generation of GPU. That is, 32 KB per multiprocessor. 
Thus the total amount of shared memory on the card
was reduced, while the number of cores increased dramatically.
Providing these cores with enough work proved to be a challenge.
The issue was resolved by moving the walker data into the register space of the SMX. 
The Kepler chipset is equipped by 512 KB of register space per SMX, which is considerably larger 
than the available shared memory. 
Thus the walker data is partitioned, and each thread stores into its registers
the coordinates for several particles of the walker. 
When a particle is displaced for a trial move, 
the holding thread releases necessary coordinates to the rest of the threads through shared memory. 
This approach allowed for considerably more resident walkers on an SMX,
giving better occupancy and increased performance. 
For the benchmarking, all K20 results were obtained with such \emph{register heavy} routines.
The routines are provided as part of the released program and may be selected  through 
a preprocessor macro. 
In addition to improved performance, the use of  the register space on the K20 allows to 
compute much larger systems, in excess of ten thousand atoms.

\subsection{Quantum Liquids package \label{sec:QLPackage}}

The CUDA routines that are described here are 
provided along with the code 
code that is used to properly launch the calculations.
The GPU code is written in the C implementation of CUDA \cite{cudamanual}, while the supporting code 
is provided in Fortran.
The resulting program is distributed
as a package through the Computer Physics Communications Program Library.
We refer to this program as the QL package.  
The QL package is a simplified version of 
a larger quantum fluids and solids package.
The simplification allows us to focus on the GPU parallelization
scheme. However, only basic functionality in terms of the
observables is included. 
The supplied package files include 
a short manual. The manual lists the file structure of the program, 
input and output files, and control variables. 
Interested readers should refer to the manual 
for setting up the calculations.
The distributed package is programmed to simulate 
\textsuperscript{4}He with 
HFD-B(HE) (known as Aziz-II)
interatomic potential  \cite{AzizII}
and 
a Jastrow wavefunction 
as described by Eqs.~(\ref{eq:Jastrow}),(\ref{eq:uFunction}),(\ref{eq:fifthPower}).
It should be relatively straightforward to adapt the package
to other bosonic systems, especially if the wavefunction remains in the Jastrow form.
The list of necessary changes is detailed in the manual.

The QL package is distributed in the form of a compressed .tar archive. 
Unpacking creates a directory called \cs{ql}, 
which contains a makefile and the makefile configuration script, 
and a number of subdirectories. 
Compilation requires the CUDA compiler nvcc and the GNU Fortran compiler.
The make command should be invoked directly in the directory \cs{ql}. For a standard Linux distribution,
executing 
\begin{verbatim}
make ql
\end{verbatim}
in the command shell
may be sufficient to compile the program.
Alternatively, users may first run the configuration script, 
and execute
\begin{verbatim}
make configure
make clean
make ql
\end{verbatim}
The configuration script tries to locate the libraries and 
necessary compilers.

Upon successful compilation, the user should  be able to change to
the subdirectory \cs{bin} and execute the program by entering
\begin{verbatim}
./ql
\end{verbatim}
in the command prompt.
The package is distributed with the  input files 
that are ready to simulate liquid \hefour\ at its experimental equilibrium density,
with 1000 atoms and 16 walkers. 
The program should finish in less than three minutes.
This sample run will execute 20 computational blocks, each block consisting of 100 analysis instances.
Each analysis is separated by 4 macroupdates.
Thus the sample run executes 128~million single-particle updates.
During the execution, several data files are produced in the same directory.
These include \cs{e.dat} with the energy values from each block,
\cs{acceptance.dat} with the average Metropolis acceptance rates,
and \cs{final\textunderscore config.dat} with the last walker coordinates. 
File \cs{kinetic.dat} contains two independent estimates for the kinetic energy.
The values in the second and third columns of this file should be statistically equal.
File \cs{store\textunderscore seed.dat} contains the last used seed of the pseudorandom sequence.
The details of the output file format are described in the provided manual, 
which can be found in the subdirectory \cs{ql/doc}.
The contents of \cs{e.dat} become as follows (not all digits are shown),

\begin{tiny}
\begin{verbatim}
      1       -5.8007    -20.9134     15.1127
      2       -5.8122    -20.9253     15.1131
      3       -5.7851    -20.9589     15.1738
      ...
     19       -5.8102    -20.9566     15.1464
     20       -5.7813    -20.9187     15.1373
\end{verbatim}
\end{tiny}
The acceptance rates in file \cs{acceptance.dat} become
\begin{tiny}
\begin{verbatim}
      1        0.3725
      2        0.3727
      3        0.3722
      ...
     19        0.3720
     20        0.3723
\end{verbatim}
\end{tiny}

It should be noted that the produced output depends
on the pseudorandom sequence used, which is compiler-dependent.
Only the averages of the computed values need to be equal between different compilers.
The distributed package uses Fortran built-in generator.
We have used the code with the GNU Fortran compiler, which implements
Marsaglia's well-known KISS generator. 

The parameters file \cs{parameters.in} specifies the execution configuration,
including the number of the computational blocks (field \cs{blockstogo}),
the number of analysis instances, or writeouts, in a block (\cs{stepsinablock}),
number of macroupdates between writeouts (variable $m$ in Fig.~\ref{fig:Workflow}; field \cs{vmcmacrosteps}),
dimensions of the simulation volume (field \cs{bulkdimarray}), parameter $b$ for the wavefunction (see Eq.(\ref{eq:fifthPower}); field \cs{jastrowb}),
particle mass in a.m.u.\ (\cs{mass}), and the amplitude multiplier for the random displacements (\cs{mtpstep}).
Changes in these parameters do not require recompilation.

Variables that require recompilation are included in the form of preprocessor macros in 
file \cs{setup.h}, which is 
located in the source directory \cs{ql/src}. 
The number of particles, walkers, threads, 
and the functionality which is performed on the GPU are set in \cs{setup.h}.
The file \cs{setup.pp} is the preprocessor header file that is used for the Fortran source files, 
and it is automatically produced by Make by stripping the comment lines from \cs{setup.h}. 
Therefore, one should never have to edit \cs{setup.pp}, which is automatically removed by 
Make after the compilation.

In order to accumulate sufficient statistics, it is often desirable to 
chain the execution of the Monte Carlo program. To do this,
one has to copy the last walker coordinates and the last pseudorandom sequence seed
into input files, by executing
\begin{verbatim}
cp final_config.dat configuration.in
cp store_seed.dat seed.in
\end{verbatim}
After the copy commands, the program may be launched again and it will continue 
from the point at which it had last finished.
Values for energy, acceptance rate, and pair distribution function  histograms will be appended
to the already existing data files.

While we have tested the program with the currently available Fermi and Kepler architectures, 
the code should be future-compatible with the upcoming generations of GPU. The Kepler-specific 
optimization  should be useful for the announced Maxwell and Volta cards, since it amounts to the extensive use of the expanded register space. 
Users can toggle such optimization on and off  with a preprocessor macro, as described in the manual.

\section{Benchmarking conditions \label{sec:Benchmarking}}

\newcommand{\ra}[1]{\renewcommand{\arraystretch}{#1}}
\begin{table*}\centering
\caption{\label{tab:Architecture}Configuration of the computers 
used for benchmarking. The top column shows the labels used to
refer to the corresponding machine in the text. 
M2090s were accessed on a BullX GPU cluster in the Barcelona Supercomputing Center. 
The 560Ti and K20 were installed on a workstation.
}

\ra{1.2}
\begin{tabular}{@{}lll>{\columncolor[gray]{.8}[6pt][6pt]}llll>{\columncolor[gray]{.8}[6pt][6pt]}llrrr@{}}\toprule&&& \textbf{560Ti} && \textbf{M2090} && \textbf{K20}
\\\midrule
\multicolumn{2}{l}{\hspace{-0.5em}Architecture} & \phantom{} & Workstation & \phantom{} &  Cluster& \phantom{} & Workstation\\[5px]
CPU  && && && && \\
\multicolumn{2}{l}{\hspace{+0.5em}  make  }                && Intel && Intel  && Intel \\
\multicolumn{2}{l}{\hspace{+0.5em}   type  }                && i5-2500 && Xeon E5649  && i5-3570 \\
\multicolumn{2}{l}{\hspace{+0.5em}  clock frequency }       && 3.3 GHz && 2.5 GHz && 3.4 GHz  \\
\multicolumn{2}{l}{\hspace{+0.5em}  memory }                &&  16 GB && 24 GB   && 12 GB   \\ 
\multicolumn{2}{l}{\hspace{+0.5em}  L3 cache }              &&  6 MB && 12 MB   && 6 MB   \\ [5px]  
GPU && && && &&  \\
\multicolumn{2}{l}{\hspace{+0.5em}   series}                && GeForce         &&  Tesla       &&   Tesla \\
\multicolumn{2}{l}{\hspace{+0.5em}   architecture}          && Fermi           &&  Fermi       &&   Kepler \\
\multicolumn{2}{l}{\hspace{+0.5em}   chipset}               && GF114           &&   T20A       &&   GK110 \\
\multicolumn{2}{l}{\hspace{+0.5em}   model}                 && GTX 560 Ti      &&  M2090       &&   K20  \\
\multicolumn{2}{l}{\hspace{+0.5em}   compute capability}    && 2.{\tiny1}      &&    2.{\tiny0}&& 3.{\tiny5} \\
\multicolumn{2}{l}{\hspace{+0.5em}   number of cores}       && 384             &&    512       &&  2496  \\
\multicolumn{2}{l}{\hspace{+0.5em}   \textsc{sm(smx)} count}&& 8               &&     16       &&  13    \\
\multicolumn{2}{l}{\hspace{+0.5em}   cores per \textsc{sm}} && 48              &&     32       && 192  \\
\multicolumn{2}{l}{\hspace{+0.5em}   clock frequency}       && 1.8~GHz         &&    1.3~GHz   &&  0.71~GHz \\
\multicolumn{2}{l}{\hspace{+0.5em}   global memory}         && 1.0~Gb          &&    6.0~Gb    &&  5.0~Gb \\
\bottomrule
\end{tabular}
\end{table*}


The reported acceleration values are ratios of times
measured for the GPU-parallelized and serial code executed on the same machine.
We have decided to compare the execution of our CPU+GPU 
code to the execution on a single-core of a CPU,
notwithstanding the fact that a GPU is usually accompanied by a powerful 
CPU with several cores.
Notice that the CPU+GPU execution also uses only a single CPU thread. 
One may argue that the acceleration 
should be compared with 
a run which utilizes 
all available CPU cores.
However, we find the single-core comparison
appropriate for several important reasons.
First of all, in practice, the calculations 
of the kind described here are so numerically demanding that they
are performed on large clusters or supercomputers.
The decision about using a GPU accelerator in such 
a case is guided by comparing 
available GPUs to 
an available allocation of core-hours on
a supercomputer.
Second, the code that uses several cores needs to 
be parallelized, and questions may arise 
regarding the efficiency of such 
parallelization. Inefficient OpenMP implementations
will artificially boost the apparent GPU acceleration numbers.
A skeptical reader may always divide our results
by the number of cores on the CPU, corresponding to an ideal OpenMP implementation,
but the reverse is not possible without careful (and distracting) performance analysis of the CPU parallelization.
Finally, single-core benchmarking is a widely accepted practice and allows for a better comparison with other works.

We note that the same optimized routines were also written for the CPU.
Thus the CPU-only program
is optimized in good faith just like the GPU+CPU program.
While the CPU was always used with a single core, automatic SIMD vectorization  for the CPU 
was used on the compiler level. SIMD instructions allow for a small level of data parallelism already on the 
single-core level of a CPU. It is possible that a full manual optimization of the SIMD parallelization
would additionally improve the CPU version of the program. However, we judged against such parallelization.

The Jastrow wavefunction that is used in this code is on purpose 
a rather simple one for its class. 
The mathematical load is small, while all the pair distances nonetheless need to be computed, 
and thus all the memory operations are still necessary, along with the pseudorandom
number generation.
In practice, one often uses trial functions that are much 
more elaborated. 
Additional arithmetic complexity of the wavefunction  
in fact improves the parallel performance.
In other words, we are testing the code with an unfavorable
trial wavefunction.

Performance tests have been carried on several machines with different Nvidia GPU models.
The computer specifications are summarized in Table~\ref{tab:Architecture}.
The Fermi-architecture 560Ti is a card from the GeForce family.
It is a popular and very affordable card with excellent heat and noise control. 
Our 560Ti was overclocked by the manufacturer.
The Tesla M2090 is a Fermi card designed specially for scientific computing. 
Tesla M2090s were accessed in a GPU cluster provided by the Barcelona Supercomputing Center.
It should be noted that the Xeon processor accompanying the M2090 has a 
slower clock than the i5 CPUs in the other machines. 
This influences the acceleration ratios 
presented later, and one should 
be aware of this when trying to compare different cards.
Finally, the Tesla K20 card
was used in an Intel i5 workstation. 
The Tesla K20 is the latest Kepler architecture card from Nvidia.
Kepler devices provide a significantly increased number of cores (in this case, 2496), 
a much larger register space, and instructions for improving global memory caching.

\section{Results\label{sec:results}}

Calculations were performed for a system representing
liquid helium at its equilibrium density $\rho=21.86$~nm\textsuperscript{3}. 
The wavefunction was of the Jastrow type~(\ref{eq:Jastrow}), (\ref{eq:uFunction}), (\ref{eq:fifthPower}),
with the parameter $b$ at optimum value $b=0.307$~nm.
The single-particle displacement was Gaussian-distributed with a root mean square
displacement of $0.5 \rho^{-1/3}$, which results 
in the  acceptance of roughly $40\%$ of trial moves.
We used the HFD-B(HE) (known as Aziz-II) interatomic interaction potential \cite{AzizII}.
The simulation box was always cubic with periodic boundary conditions.
The cutoff distance for the potential and for the correlation factors (length $L/2$ in Eq.~(\ref{eq:uFunction})) was set at half the box size.
Calculations were performed on three different cards, as detailed in Section~\ref{sec:Benchmarking} and specified in Table~\ref{tab:Architecture}.

\subsection{Load masking \label{sec:PairDistributionMasking}}

\begin{figure*}
\makebox[1\columnwidth][c]{ %
\begin{centering} %
\includegraphics[width=0.6\columnwidth]{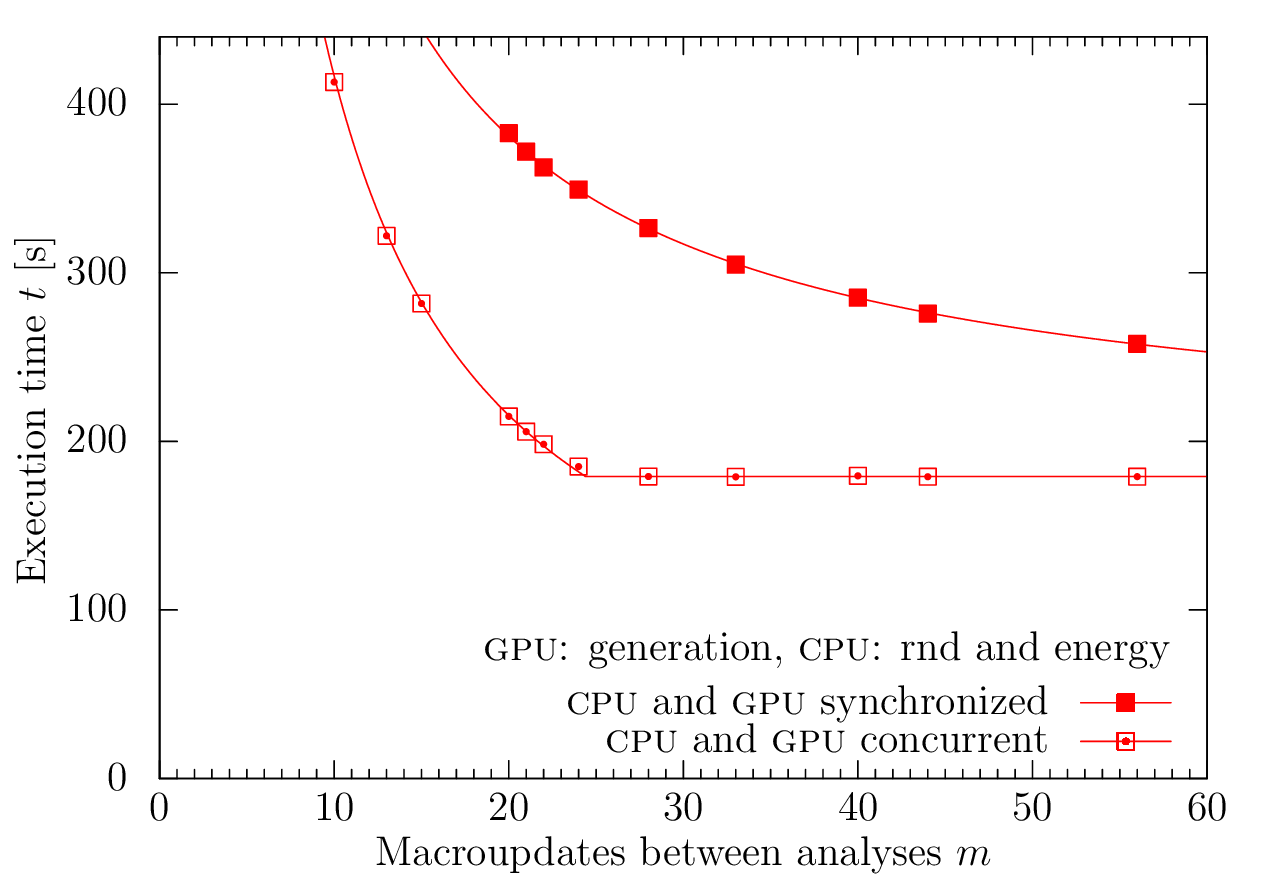}
\hspace*{1em}
\includegraphics[width=0.6\columnwidth]{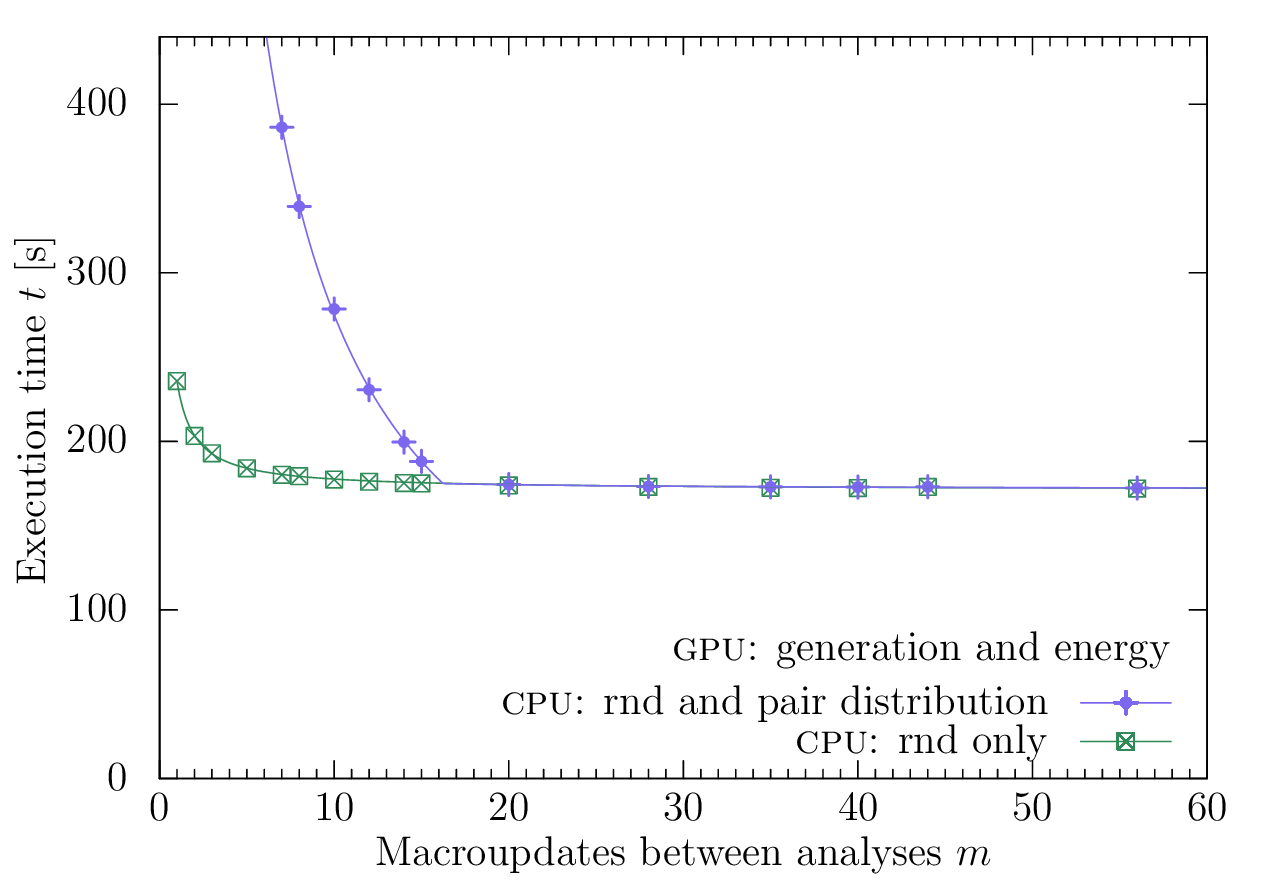}
\end{centering} %
}
\caption{
\label{fig:Masking}
(Color online)
Load masking in a calculation with $N=512$ particles and eight independent Markov chains on a GTX560Ti.
The plots show the total execution time in seconds vs the number of macroupdates between analyses.
The total number of macroupdates produced by the GPU for each chain was
the same for each run and equal to 92400 (broken into 10 blocks).
Every $m$ macroupdates, the configurations were written out
to the host RAM and some analysis was performed.
The vertical axis shows the total execution time in seconds.
All lines are fits to the corresponding regions in the form $a+b/m$.
\emph{Left:} Configurations are produced by the GPU, energy is computed 
on the CPU. Filled squares show execution times for execution with synchronized kernel (that is, CPU and GPU sides execute sequentially).
Open squares are times for asynchronous (concurrent) execution with load
masking as described in the text.
\emph{Right:}
The GPU produces configurations and computes the energy.
Green crossed boxes: CPU receives the configurations
but does not perform any analysis.
Blue crosses: The CPU receives the configurations and 
computes the pair distribution function. 
In both cases, the execution is asynchronous.
For $m>20$, the calculation
of the pair distribution function is  completely masked.
}
\end{figure*}

First we demonstrate masking of the CPU load
by the GPU execution as described in Section~\ref{sec:LoadMasking}.
The calculations were made on the 560Ti card for eight independent Markov
chains, one walker per SM. 
The GPU was used to execute the Box-Muller algorithm and then to generate new configurations, 
while the CPU-side generated 
the raw random numbers and performed energy calculations
for configurations separated by a variable number of macroupdates $m$.
The total number of macroupdates was kept fixed
while the number of analysis 
instances was varied.
Thus the work performed by the  GPU was not varied,
and  the  needed number of random numbers was also fixed.
Given the total number of macroupdates is $s$ (in this case, 10 blocks of 9240 macroupdates),
the number of writeouts and energy analyses is $w=s/m$. Thus the CPU time should be 
of the form $a+b/m$, where the constant $b$ includes the cost of computing the energy.

The results for the execution time are shown in the left panel of Fig.~\ref{fig:Masking} with open boxes. As can be seen, for small 
number of macroupdates between measurements
$m$, the execution time is indeed very sensitive to $m$. 
In this region, the CPU time dominates, as it is larger than the GPU time. 
However, for $m>25$, the CPU time becomes ``undetectable'' 
as it decreases below the GPU load. For a system with large autocorrelation
times, such masking would allow to keep all the analysis calculations without parallelizing them for the GPU,
with no performance consequences. 
Moreover, any additional overhead on the CPU is also masked, in accordance
with the modified Amdahl law of Eqs.~(\ref{eq:modifiedAmdahlFull})--(\ref{eq:modifiedAmdahlMasked}).
Notice also that the GPU-dominated region in Fig.~\ref{fig:Masking}
is perfectly flat, yet the number of configuration writeouts from the GPU to host RAM is decreasing as $m^{-1}$. 
The GPU is able to continue its work before the memory transaction is complete, and the 
cost of configuration writeouts is masked to the level of being negligible.

The liquid helium system that we consider has short autocorrelation times regardless of the system size.
For optimum performance, the energy calculations had to 
be moved to the the GPU as well. The right panel of Fig.~\ref{fig:Masking} shows the resulting
execution times (with green crossed boxes). Now the GPU time consists of a constant generation 
time and the time for computing the energy which scales as $m^{-1}$. 
The same figure shows masking of another 
operator (with blue crosses). In that case, configurations and the energies are 
computed on the GPU, while the CPU performs calculations of the pair distribution 
function, a quantity of interest for such systems. Again there is a 
\emph{masking threshold} in $m$ above which the additional  calculation comes at no cost.

\subsection{Generation speed \label{sec:generationspeed}}

\begin{figure*}
\makebox[1\columnwidth][c]{ %
\begin{centering} %
\includegraphics[width=0.6\columnwidth]{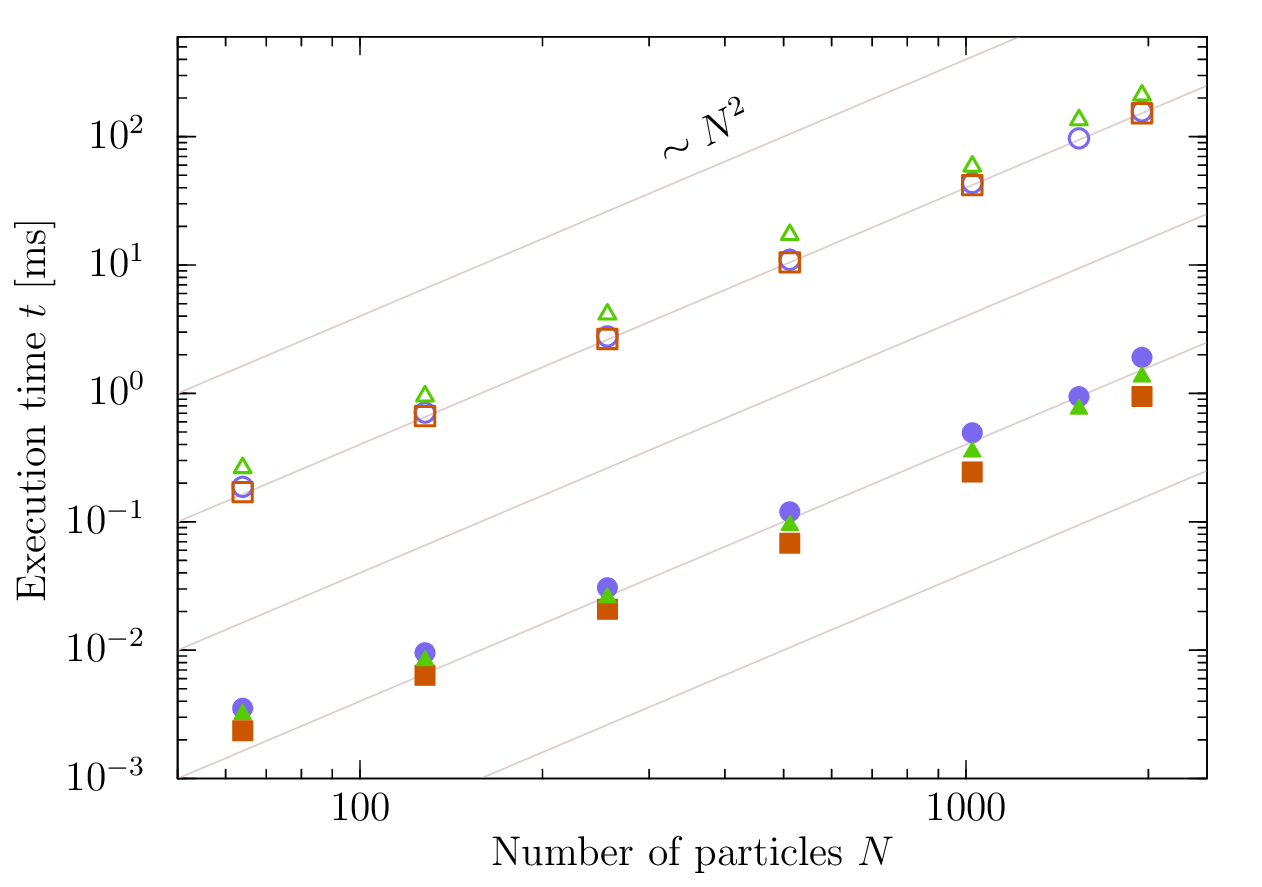}
\hspace*{1em}
\includegraphics[width=0.6\columnwidth]{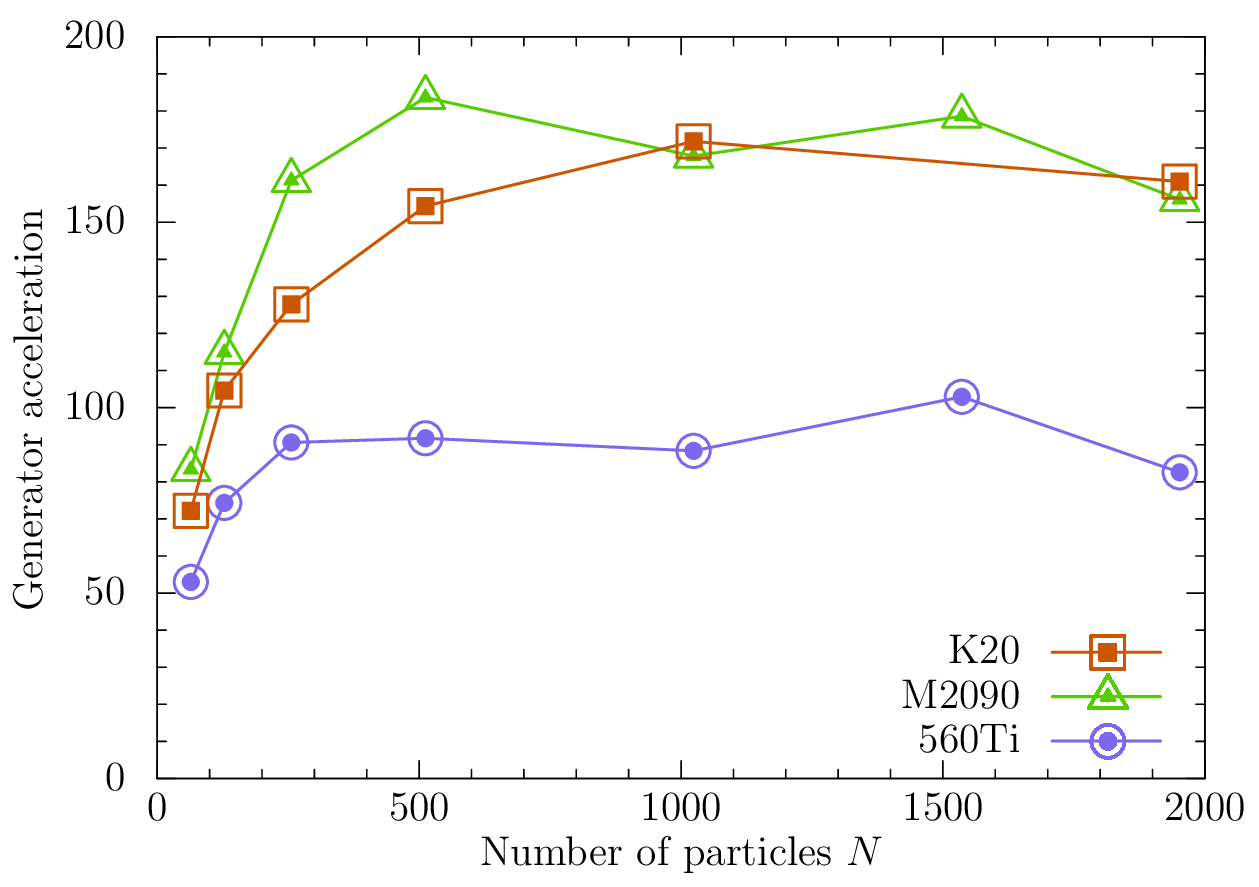}
\end{centering} %
}
\caption{
\label{fig:generatorAcceleration} (Color online) 
Generation of new configurations by GPU 
compared to a single-core generation on CPU, for a varying number of particles $N$.
The generation time is shown per macroupdate per Markov chain.
Blue bullets: 560Ti; green triangles: M2090; red squares: K20.
GPU details are given in Table~\ref{tab:Architecture}.
\emph{Left:} Absolute time, in milliseconds, necessary to perform a macroupdate.
Open symbols correspond to the CPU-only execution with a single core. Closed symbols:
GPU-accelerated generation. 
Diagonal lines show $\sim N^{2}$ behavior.
\emph{Right:} Resulting acceleration ratios, i.e. ratios of generation times with or without GPU acceleration for each system.
Notice that generation times are in fact lowest for the K20 GPU}
\end{figure*}

In this section we look at the speed in which the GPU can generate new configurations.
This is especially important for properties with large correlation times.
The GPU was only used to generate new configurations, for as many independent 
Markov chains as was found optimal to achieve the maximal throughput.
In practice this meant up to 208 walkers for the smallest $N=64$ systems on Kepler K20 (16 walkers per SMX).
The optimal number of resident walkers per SM depends on the model of the GPU.
For example, for the largest system shown in Fig.~\ref{fig:generatorAcceleration} ($N=1952$), Fermi M2090 was optimally ran with
two walkers per SM (32 walkers in all), while Kepler K20 could be efficiently loaded with
up to five walkers per SMX (65 walkers per card).
For CPU-only execution, one chain is optimal to make the best use of the L1 cache.
The results are shown in Fig.~\ref{fig:generatorAcceleration} in 
the form of times necessary to perform a macroupdate and the acceleration ratio 
for each GPU model.

As can be seen, the acceleration is remarkable. 
Both for GPU and CPU, the macroupdate generation time scales
with the system size as $N^2$, and the resulting acceleration is roughly constant 
and equal to $\times85$ for the 560Ti card, and over $\times150$ for M2090 and K20.
However, performance is reduced for small systems, where 
random number generation becomes a limiting factor.
The random number generation scales as $\mathcal{O}(N)$ and 
is masked for systems with more than $256$ particles, as discussed 
in Section~\ref{sec:LoadMasking}.

Notice that these results apply to the case when the system has reached equilibrium
and one is interested in determining the observables with best accuracy, as described 
in Section \ref{sec:LoadMasking}. For this, one needs to generate maximum overall
number of new configurations for all Markov chains combined. For the equilibration stage,
one would use a smaller number of walkers, perhaps equal to the number of 
multiprocessors on the card (between 8 and 16, see Table \ref{tab:Architecture}). The best
acceleration for the progression of each individual walker that we observed
did not exceed $\times12$ when the walkers are served by individual multiprocessors. 
When dealing with extremely slowly equilibrating systems, one may have to resort 
to updating a single walker with the cooperating kernels, when a single configuration is 
updated with the entire GPU, as mentioned in Section~\ref{sec:CUDA}. 

\subsection{Full acceleration \label{sec:energybenchmarking}}

\begin{figure*}
\makebox[1\columnwidth][c]{ %
\begin{centering} %
\includegraphics[width=0.6\columnwidth]{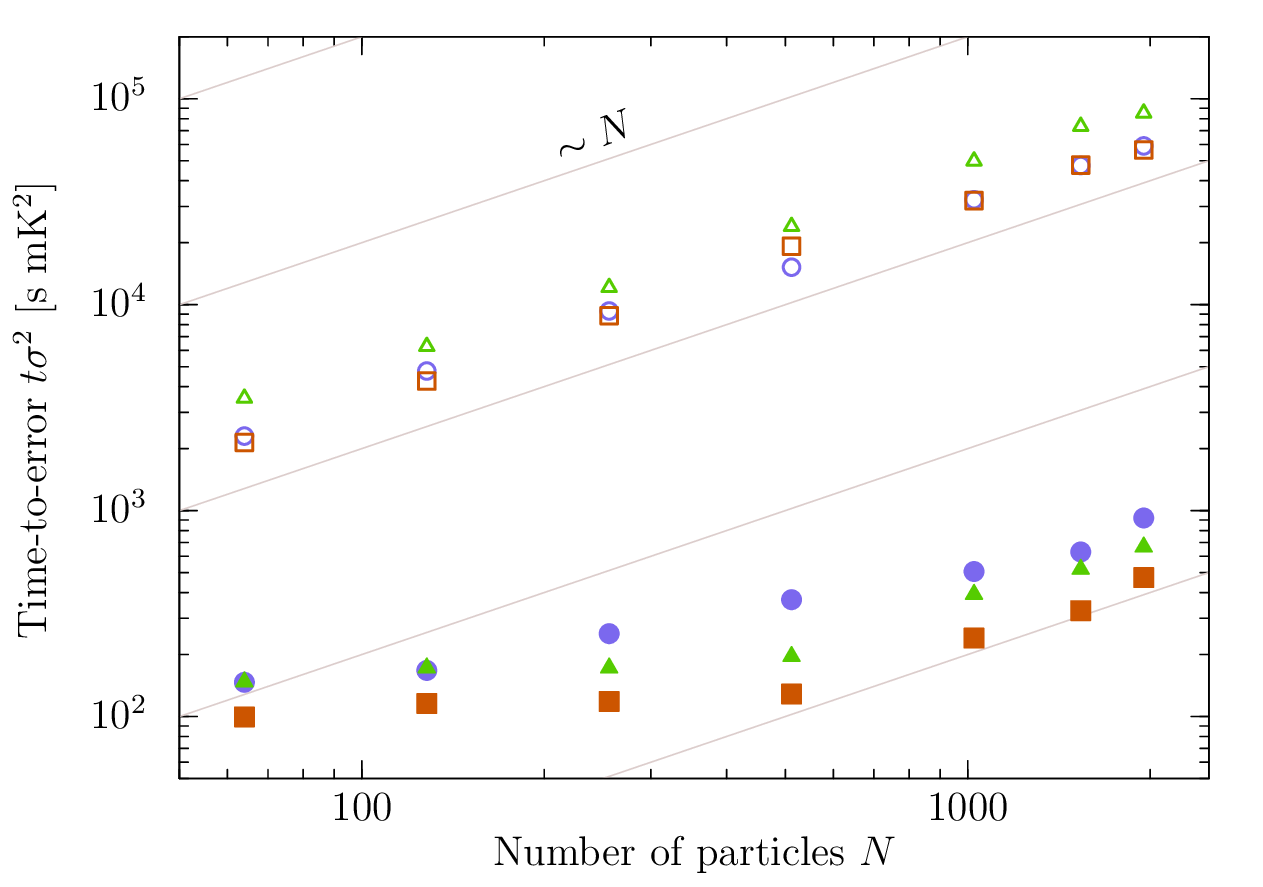}
\hspace*{1em}
\includegraphics[width=0.6\columnwidth]{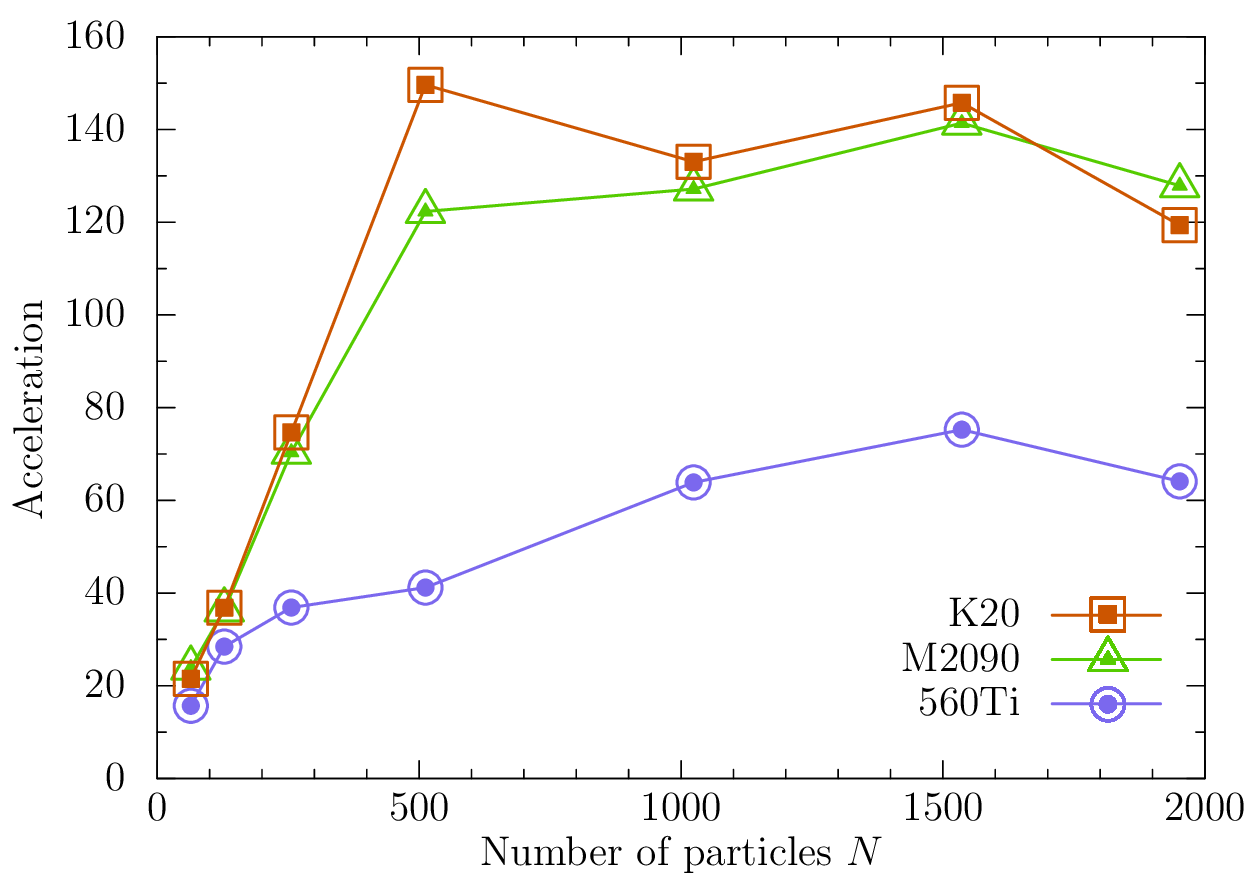}
\end{centering} %
}
%
\caption{\label{fig:totalAcceleration} (Color online) 
The speed of computing the energy to a given accuracy as a function of the number of particles $N$.
Blue bullets: 560Ti; green triangles: M2090; red squares: K20.
\emph{Left:} Time, in seconds, necessary to compute the energy with the error of 1mK per atom.
Open symbols correspond to a single-core CPU execution, and closed symbols show 
GPU-accelerated execution. 
Lines show $\sim N$ behavior. 
\emph{Right:} Resulting acceleration, i.e. ratios of execution times measured with or without GPU acceleration (show on left panel) for each system.
}
\end{figure*}

Speed gains should ultimately be judged from the time that is necessary to compute the observables.
We focus on the energy, while other properties have a similar computational complexity.
To determine the acceleration, we compared execution 
times that are necessary to reach a given error level in the computed energy per particle.
Both parallel and serial programs were ran with their respective optimal number of macroupdates between analyses, determined empirically.
Optimal execution is understood as such that leads to the smallest final uncertainty in the computed value of the observable, 
as described in Section~\ref{sec:LoadMasking}.
Because the GPU-accelerated and CPU-bound codes have
different ratios of analysis-to-generation costs $\alpha$, their optimal number of macroupdates between analysis $m$
are different (see Eqs.(\ref{eq:optimalCPU}--\ref{eq:optimalGPU})).
Benchmarking conditions are detailed in Section~\ref{sec:Benchmarking}.
Only one core was used on the CPU-side.
For the accelerated code, the raw random numbers were generated on the CPU-side,
while the GPU performed the Box-Muller transformation,
generated new configurations and 
computed the energy values. 
The number of walkers was 
chosen such that it provided 
the best occupancy for each number of particles $N$.
The optimal number of walkers varies between the cards. 
Best results for K20 were achieved with 8 to 12 walkers per multiprocessor (between 104 and 156 walkers in total),
even for the largest systems shown in Fig.~\ref{fig:totalAcceleration}.

The benchmarking results are shown in Fig.~\ref{fig:totalAcceleration}.
The left panel shows the time, in seconds, that is necessary to reach the 
errorbar of 1~mK. In agreement with the central limit theorem, 
the product of the execution time $t^{\text{exec}}$ and the square of the resulting
error $\sigma$ is an invariant,
\[
t^{\text{exec}} \sigma^2 = \operatorname{const}.
\]
Both for the CPU-only and GPU-accelerated execution
this constant scales with the first order of the number of particles $N$, 
due to self-averaging.
This relation can be used to  estimate the necessary execution 
time  for a desired accuracy level.
At low particle numbers, acceleration is limited by the random numbers generation.
As should be expected, this limitation is more severe for the higher-performance Kepler GPUs.
For all GPUs, we observe linear scaling of the time-to-error for systems with more than 512 particles.
For a 512-atom system, a 1~mK errorbar is reached in just over two minutes on Kepler K20 ($t\sigma^2=130\,\text{s}\cdot\text{mK}^2$), 
and in under four minutes on M2090 ($t\sigma^2=200\,\text{s}\cdot\text{mK}^2$).
It is worth pointing out that this corresponds to a $2\cdot 10^{-4}$ relative accuracy, 
more than satisfactory for most applications.

The right panel of Fig.~\ref{fig:totalAcceleration} shows
the resulting acceleration for each card.
The values degrade for small particle number,
as commented above. For systems with more than 500 atoms, 
acceleration ratios remain relatively constant.
For the 560Ti, the value equals to $\times$50, 
and for the Tesla M2090 and K20, acceleration exceeds a factor of $\times$120.
K20 was faster than M2090 for all system sizes.

\subsection{About interpreting the acceleration values\label{sec:performancewarning}}

Readers should be cautious when interpreting the acceleration ratios
presented on the right panels 
of Figs.~\ref{fig:generatorAcceleration} and \ref{fig:totalAcceleration}.
Our benchmarks show acceleration by a factor of over $\times$150 for generating
new configurations and over $\times$120 for the combined generation and 
energy calculations. As explained in Section~\ref{sec:LoadMasking}, these
were obtained by comparison with single-threaded execution on the CPU. 
Modern multithreaded CPUs, however, have up to eight cores and are capable of supporting as many as 16 threads. 
Thus when comparing with the computational capability of an \emph{entire}
CPU, presuming nearly-ideal parallelization, 
one should remember to divide by the supported number of threads.
Misunderstandings of this sort often plague the interpretation of the GPU accelerator
capabilities.

The true ``benchmark'' for a program of this kind is its ability to compute observables.
For quantum Monte Carlo calculations, this amounts to the ability to reach a desired level of statistical accuracy.
We have thus focused on the generation of the  Markov chain in general, and also used the energy as a reference observable.
The performance for both of these functions can be clearly characterized in absolute terms. 
For the Markov chain generation, one can find the rate at which new configurations can be produced,
and for the energy calculation, one can record the time-to-error as explained above.
This information is provided in the left panels of Figs.~\ref{fig:generatorAcceleration} and \ref{fig:totalAcceleration}.
Such ``observables performance'' can be readily compared between different programs and even methods.

\subsection{Notes on the Kepler K20 performance}
 
It is worth pointing out that the K20 was the best performing card.
The execution times, shown in the left panels 
of Figs.~\ref{fig:generatorAcceleration} and~\ref{fig:totalAcceleration}, 
are smaller for the K20 than for the other two cards.
The acceleration, shown in the right panels of the two figures, seems larger for the M2090
because the individual CPU cores of the machine with the M2090 where slower, 
thus slightly inflating the acceleration. In fact, the K20 was as much as one-third
faster than the M2090.
This is remarkable given the already strong performance of the M2090.
The improvement was achieved by using the register-heavy 
routines as described in Section~\ref{sec:Implementation}.
The extraordinarily large amount of register space on Kepler chips
allowed to simultaneously process a larger number of walkers and thus keep
the card occupancy high.

The second advantage of the large register space of the K20 is that it is capable of handling 
much larger systems when using the register-heavy routines, as 
described in Section~\ref{sec:Implementation}.
The scaling behaviors shown in Figs.~\ref{fig:generatorAcceleration}~and~\ref{fig:totalAcceleration} 
are preserved even for the largest systems. 
This is despite the fact that for large systems, one is limited to 
a single walker per multiprocessor.
For example, generating one macroupdate for a 40960-particle calculation takes 418~ms 
(this number takes into account the fact that there were 13 walkers in total).
The energy-computing kernel uses extra memory to
store wavefunction derivatives; this kernel was used with up to 10240 particles.
The ability to treat such large systems opens an access to a range of new applications.

\section{Conclusion}

We have developed a remarkably efficient parallelization of 
bosonic variational Monte Carlo for graphical processing units. 
Up to two thousand particles may be treated with an Nvidia Fermi GPU, and up to
ten thousand with Nvidia Kepler cards. 
The GPU exhibits excellent speed results, which we measure as 
the time necessary to reach a level of statistical uncertainty in observables. 
The good acceleration is mainly due to two 
developments. First and foremost, the execution was organized to allow for 
shared memory persistence. 
The thread blocks of the GPU run uninterrupted
for multiple steps, maintaining their walker information in shared memory or even in
the register space. 
Second, the produced configurations are written out by the GPU to host memory as the calculation progresses.
The analysis and observable calculations can occur both on the GPU and on the CPU.
The analysis on the CPU occurs simultaneously with the GPU generating the next configuration.
This masks the CPU load and allows the developer to focus on improving the parallel fraction of the generating GPU kernel. 
The asynchronous execution results in a modified version of the Amdahl law.
Existing CPU-based routines may be readily applied to the 
generated configurations, easing the transition to the GPU-parallel execution.
The code is distributed in a form of a package that is ready to simulate liquid \textsuperscript{4}He.
With minor modifications, it can be rendered to apply to a range of bosonic problems.
 

\section*{Acknowledgements}

Author would like to thank the Barcelona Supercomputing Center 
(The Spanish National Supercomputing Center -- Centro Nacional de Supercomputaci\'on)
for the provided GPU facilities.












\end{document}